\documentclass[superscriptaddress,
               floatfix,
               longbibliography, 
               showkeys,apl,twocolumn,nofootinbib]{revtex4-2}

\usepackage{graphicx} 
\usepackage{bm}	
\usepackage{color}                     
\usepackage{xcolor}
\usepackage{epsfig}
\usepackage{amsmath} 
\usepackage{amssymb} 
\usepackage{float}
\usepackage{mathtools}
\usepackage{dsfont}
\usepackage{dsfont}
\usepackage{xparse}
\usepackage{hyperref}
\usepackage{times}
\usepackage[normalem]{ulem}
\usepackage{sidecap}
\usepackage{appendix} 
\sidecaptionvpos{figure}{t}
\usepackage{comment}
\usepackage{physics}
\usepackage{float}
\usepackage{xfrac}

\usepackage{svg}
\usepackage{multirow}
\usepackage{algpseudocode}
\usepackage{tikz,lipsum,lmodern}
\usepackage[most]{tcolorbox}
\usepackage{simpler-wick}
\usepackage{caption}
\usepackage{subcaption}
\usepackage{ragged2e}
\usepackage{graphics}

\definecolor{darkgreen}{rgb}{0.0, 0.5, 0.0}

\newcommand{\be}{\begin{equation}}
\newcommand{\ee}{\end{equation}}
\newcommand{\bd}{\begin{displaymath}}
\newcommand{\ed}{\end{displaymath}}
\newcommand{\BE}{\begin{eqnarray}}
\newcommand{\EE}{\end{eqnarray}}

\def\c#1{\textcolor{blue}{#1}}

\begin{document}

\title{The Renormalized Yukawa Hamiltonian: Spectrum, Parton Distribution Functions, and Resource Estimates for Quantum Simulation}

\date{\today} 

\author{Carter M. Gustin}
\email{carter.gustin@tufts.edu}
\affiliation{Department of Physics and Astronomy, Tufts University, Medford, MA 02155, USA}
\author{Kamil Serafin}
\affiliation{Department of Physics and Astronomy, Tufts University, Medford, MA 02155, USA}
\author{William A. Simon}
\affiliation{Department of Physics and Astronomy, Tufts University, Medford, MA 02155, USA}
\author{Alexis Ralli}
\affiliation{Department of Physics and Astronomy, Tufts University, Medford, MA 02155, USA}
\author{Gary R. Goldstein}
\affiliation{Department of Physics and Astronomy, Tufts University, Medford, MA 02155, USA}
\author{Peter J. Love}
\affiliation{Department of Physics and Astronomy, Tufts University, Medford, MA 02155, USA}
\affiliation{Brookhaven National Laboratory, Upton, NY 11973, USA}

\begin{abstract}
We apply the Renormalization Group Procedure for Effective Particles (RGPEP) to the front form Yukawa Hamiltonian, yielding a renormalized (effective) Hamiltonian, accurate up to second order in the coupling strength.
Subsequently, we examine the spectrum and parton distribution functions produced by the renormalized Hamiltonian, and show that the addition of counterterms leads to finite results.
 Resource estimates for quantum simulation are calculated for a single `Ladder Operator Block Encoding' (LOBE), and show that the cost to block encode the renormalized Hamiltonian is comparable to block encoding the bare Hamiltonian.
    
\end{abstract} 

\maketitle

\section{Introduction}


A central problem in particle and nuclear physics is calculating bound state mass spectra and hadronic structure~\cite{PhysRev.84.350}. Classical calculations of hadronic properties are dominated by lattice QCD~\cite{Lin:2015dga}, first proposed by Wilson~\cite{PhysRevD.10.2445}. The mass spectra and structure functions of hadrons are the targets of many lattice QCD collaborations~\cite{Lin:2023kxn, davoudi2022reportsnowmass2021topical, kronfeld2022latticeqcdparticlephysics, Davoudi_2021}.
 
An alternative framework to computing hadronic structure, coined the `light-front' approach, was derived from a 1949 paper by Dirac~\cite{dirac1949}. In the light-front approach, the Lagrangian is transformed to a Hamiltonian in light-front coordinates via the energy-momentum tensor~\cite{dirac1949}. This leads to a Hamiltonian eigenvalue problem that is similar to the non-relativistic Schr{\"o}dinger equation.

While one can formulate the Hamiltonian eigenvalue problem in equal time coordinates there are two main advantages of using the front form. First, the mass-shell dispersion relation implies that positive longitudinal front form momentum corresponds to positive front form energy. A consequence of this is that for cutoff theories with no zero modes, the light front vacuum is trivial, as opposed to the complicated vacuum in instant time. Second, the solutions to the eigenvalue equation are boost-invariant~\cite{brodsky1998quantum}. Boost-invariant eigenvalues imply they are the same regardless of the choices of longitudinal and transverse bound state momentum. 


The first numerical technique to solve bound state problems in light-front field theory was called `Discretized Lightcone Quantization' (DLCQ)~\cite{PhysRevD.32.1993, pauli1985discretized}. In DLCQ, the theory is quantized in a box, and wavefunctions are represented in a plane-wave basis of momentum orbitals. An extension to DLCQ, called `Basis Light-front Quantization' (BLFQ)~\cite{PhysRevC.81.035205} uses symmetry-adapted basis functions (analogous to orbital basis functions in quantum chemistry).
Results from classical supercomputer calculations of hadronic properties using BLFQ can be found in, e.g., Refs.~\cite{PhysRevC.81.035205,PhysRevC.99.035206,PhysRevC.102.055207,PhysRevD.96.016022,PhysRevD.91.105009}. The results, in particular in~\cite{PhysRevC.99.035206,PhysRevC.102.055207}, show agreement with experimental data published by the Particle Data Group~\cite{PDG}. 

The simulation of quantum systems has been a prospective application for quantum computers since its original conception in~\cite{feynman_simulating, feynman_qc}, with the first quantum simulation algorithms developed in~\cite{lloyd1996universal, Meyer_1996, Boghosian_1997, Abrams_1999}. 
Quantum simulation includes a broad array of computational tasks, including both the estimation of static properties, and the simulation of quantum dynamics~\cite{McArdle}.
Today, quantum simulation is dominated by Trotterization~\cite{suzuki1976generalized, hatano2005finding, lie1893theorie, trotter1959product} and block-encoding~\cite{lin, poulin, qubitization} algorithms.
Many applications of quantum simulation have been explored in fields such as quantum chemistry and high energy/nuclear physics~\cite{aspuru2005simulated, wiese2014towards, Cao_2019, PRXQuantum.4.027001}. 

Quantum simulation of quantum field theories was first investigated in detail by Jordan, Lee, and Preskill in Ref.~\cite{jordan-lee-preskill}. They employ a discrete space lattice (analogous to classical lattice QCD calculations) in which the fields can be represented at each lattice point by a finite number of qubits. Many works utilize this approach~\cite{perez2021determining, echevarria2021quantum, mueller2020deeply, byrnes2006simulating, atas20212}. For a review, see Ref.~\cite{savage2025quantumsimulationsfundamentalphysics}. 

A quantum simulation framework for quantum field theories based on the light front was proposed in~\cite{PhysRevA.105.032418, Kreshchuk:2020kcz,
Kreshchuk:2020aiq}. 
The light-front approach yields quantum field theories written in the language of quantum many-body problems~\cite{PhysRevA.105.032418}, resembling quantum chemistry, which has been well-studied as an application on quantum computers~\cite{McArdle}.
The preexisting literature on quantum chemistry is an advantage of the light-front approach as a simulation method for quantum field theories~\cite{WILSON199082, qian2022solving}. 

In this work, we study quantum simulation of $1+1$D Yukawa theory in light-front coordinates, following the  work of~\cite{pauli1985discretized, PhysRevA.105.032418}. Yukawa theory, proposed in 1935 by Hideki Yukawa~\cite{yukawa1935}, is used as a simple relativistic model for nuclear physics~\cite{Machleidt:2022kqp}.
While not a fundamental theory, the Yukawa model was a historical precursor to QCD. Both Yukawa theory and QCD require strong numerical tools to compute observables, and so Yukawa theory is a useful testbed for new techniques before tackling the many issues arising in gauge theories such as QCD.

In~\cite{pauli1985discretized, PhysRevA.105.032418} classical and quantum computations of the canonical light front Yukawa Hamiltonian were performed. In this paper we perform renormalization up to second order in the coupling strength using the Renormalization Group Procedure for Effective Particles (RGPEP)~\cite{GLAZEK2000175, glazek2012perturbativeformulaerelativisticinteractions}. RGPEP is an extension of the similarity renormalization group~\cite{PhysRevD.48.5863, szpigel2000similarity}. RGPEP is a systematic method to compute perturbatively, order by order in the coupling strength, an \textit{effective (renormalized)} Hamiltonian, free of divergences, leading to a convergent mass spectrum. While perturbation theory has been used to perform renormalization in light-front field theory~\cite{brodsky1998quantum}, RGPEP accomplishes this by a unitary transformation to the canonical Hamiltonian, leading to softer ``effective" vertices. These effective vertices are exponentially suppressed compared to their bare counterparts. Divergences in the effective Hamiltonian are removed by the addition of counterterms in the bare Hamiltonian.

RGPEP has been used in numerical calculations yielding results consistent with experiment. In particular, Refs.~\cite{Serafin:2018aih, Gomez-Rocha:2023jfr} study masses of heavy-flavor baryons while Ref.~\cite{Glazek:2017rwe} studies heavy-flavor mesons. Ref.~\cite{Kuang:2022vdy} calculates masses of all-charm tetraquarks.
Most recently,~\cite{serafin2024dynamics} studies the dynamics of bound states of heavy quarks.  The previously cited works were all done in particular sectors of the QCD Hamiltonian. Refs.~\cite{Allen_1998, Kylin_1999} studied bosonic $\phi^3$ theories, while Ref.~\cite{Allen_2000} studied glueballs. These were the first examples of relativistic field theory models where the full effective Hamiltonian is calculated using RGPEP.

The paper is organized as follows. 
In Section~\ref{sec:yukawa-theory}, the canonical Yukawa Lagrangian and light-front coordinates are introduced, and the canonical Hamiltonian in momentum space is given.
In Section~\ref{sec:dlcq}, a method for discretizing the Hamiltonian is described.
The discretized Hamiltonians, written as sums of products of creation and annihilation operators, serve as the input for quantum simulation.
In Section~\ref{sec:rgpep}, the Renormalization Group Procedure for Effective Particles is described, and the explicit effective Hamiltonian up to $\mathcal{O}\left(g^2\right)$ is given. 
In Section~\ref{sec:spectrum}, the numerical mass spectrum is shown for both the bare and renormalized theories.
In Section~\ref{sec:pdfs}, the parton distribution functions, describing the momentum distributions in bound states, are computed.
In Section~\ref{be-metrics}, the quantum resources needed for quantum simulation of the bare and renormalized Hamiltonians are calculated.
We finish in Section~\ref{sec:conclusions}, which summarizes the results and discusses potential future directions of study.
\section{light-front Yukawa Theory}
\label{sec:yukawa-theory}
Motivated by~\cite{PhysRevA.105.032418} and following Ref.~\cite{serafin-yukawa}, the Yukawa model is described by the Lagrangian
\begin{equation}
    \label{eq:yukawa-lagrangian}
    \mathcal{L}_{\rm{Y}} = \bar \psi \left(i\gamma^\mu \partial_\mu - m \right)\psi + \frac{1}{2}\partial_\mu \phi \partial^\mu \phi - \frac{1}{2}\mu^2\phi^2 - g\bar \psi \psi \phi,
\end{equation}
where $m$ is the fermion mass parameter, $\mu$ is the boson mass parameter, and $g$ is the coupling strength between the fields. 
$\phi$ is a scalar bosonic field mediating interactions between $\psi$, the fermionic field.

$1+1D$ light-front (front form) coordinates~\cite{dirac1949} are defined by the transformation from instant (equal) time coordinates $x^\mu = \left(x^0, x^3 \right)$ to \textit{light-front coordinates} via:
\begin{equation}
    \begin{cases}
    x^+ \equiv x^0 + x^3 \\
    x^- \equiv x^0 - x^3
\end{cases}
\end{equation}
where $x^+$ is taken to be the light-front ``time'' coordinate, while $x^-$ is taken as the light-front ``space'' coordinate.

The light-front two-momentum in $1+1D$ is given by $p^\mu = (p^+, p^-)$, where $p^+ = p^0 + p^3$ is conjugate to $x^-$ and therefore is taken to be the light-front (longitudinal) ``momentum'', while $p^- = p^0 - p^3$ is conjugate to $x^+$ and therefore is taken to be the light-front ``energy''. Throughout, we use captial $P^\mu$ to refer to the two-momentum carried by a collection of particles, while lower-case $p^\mu$ refers to the two-momentum carried by a single particle. The ``energy'' $P^-$ generates translations in light-front time, $x^+$.

The Hamiltonian density, $\mathcal{H}$, is  $\frac12 T^{+-}$, where $T^{\mu \nu}$ is the energy-momentum tensor of the theory.
The Hamiltonian is the spatial integral of the corresponding density. 
In 1 + 1D light-front coordinates, the Hamiltonian is:
\begin{equation}
    \label{eq:H}
    H = \int dx^- \mathcal{H}.
\end{equation}
This gives the so-called canonical Hamiltonian, which is the input to any subsequent renormalization group procedure.

\begin{figure}
  \begin{subfigure}{0.45\columnwidth}
  \includegraphics[width=\textwidth]{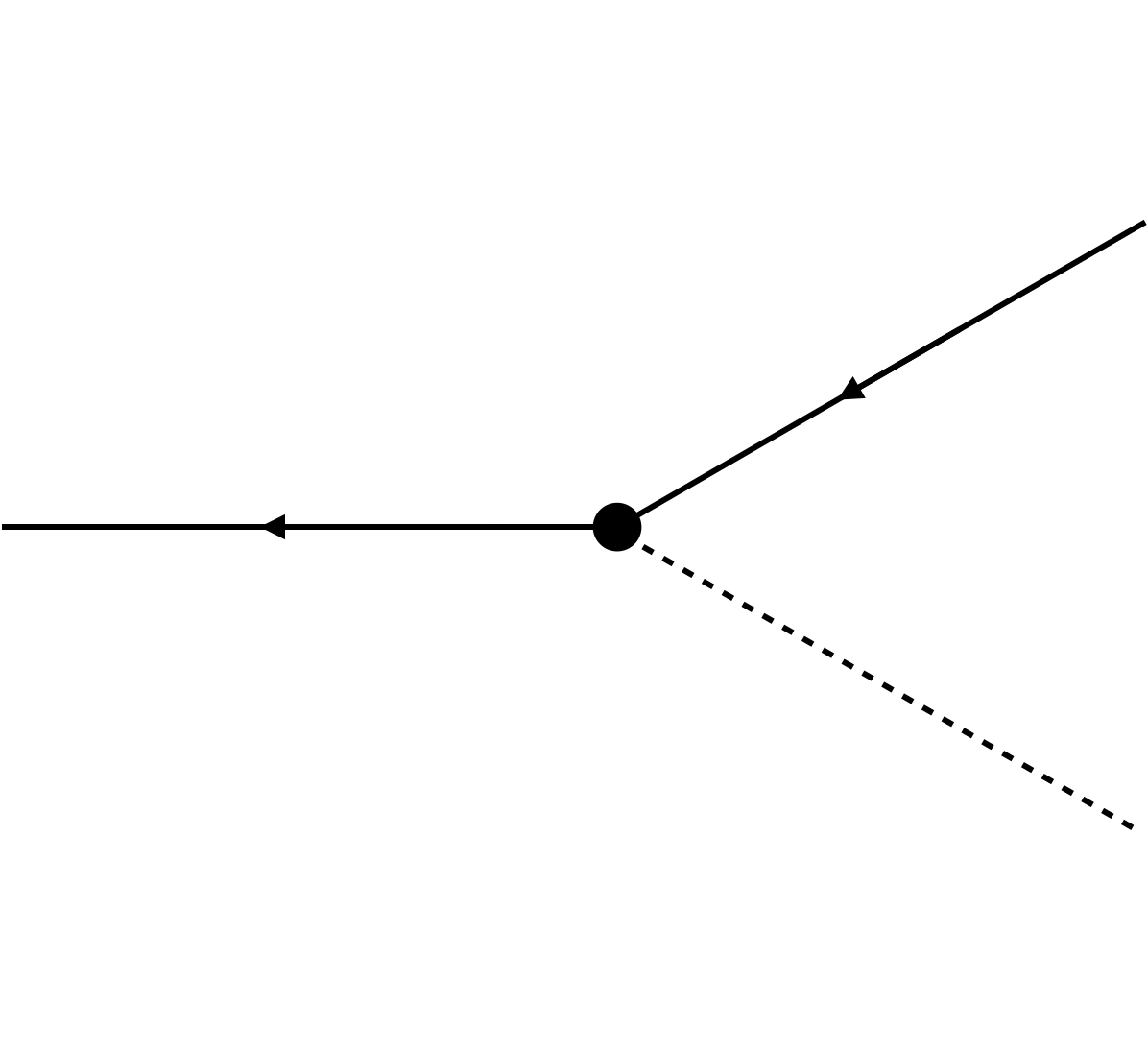}
  \caption{$H_{\bar \psi \psi \phi}$ Example Diagram}
  \end{subfigure}
  \hfill
  \begin{subfigure}{0.40\columnwidth}
  \includegraphics[width=\textwidth]{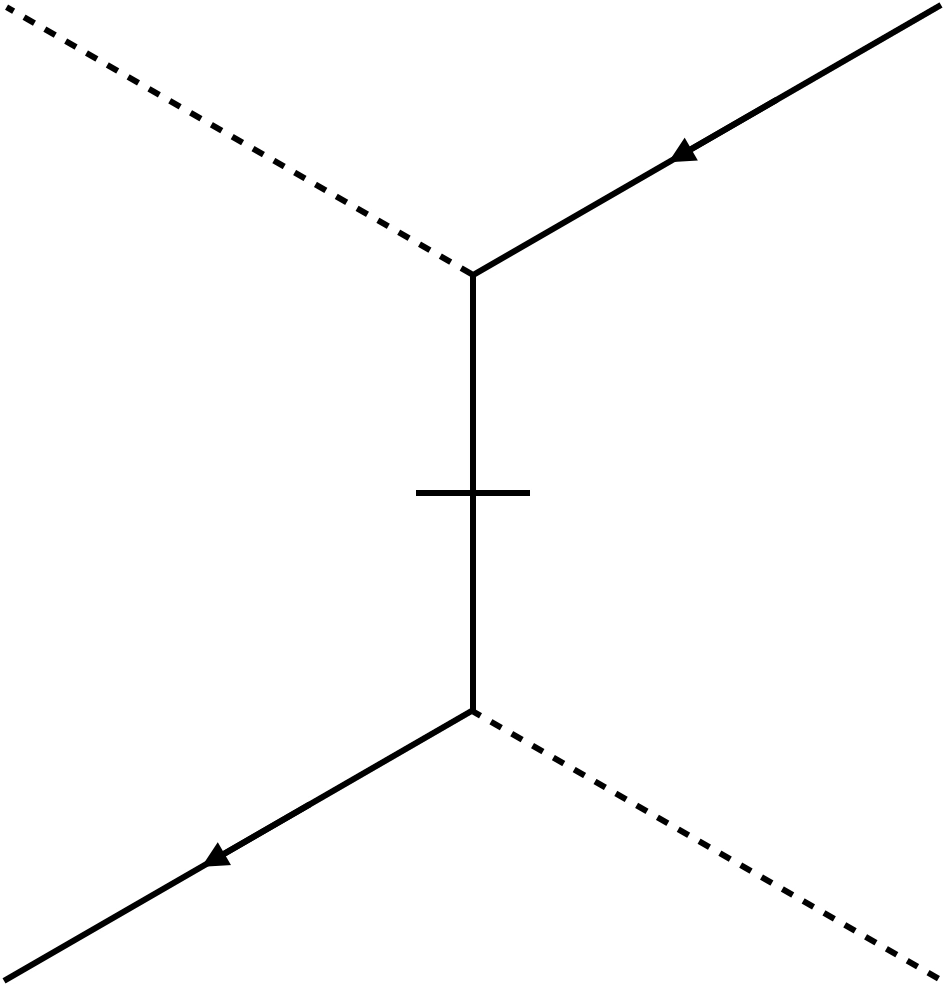}
  \caption{$H_{\bar \psi \phi \phi \psi}$ Example Diagram} 
  \end{subfigure} 
  \caption{\justifying \textbf{Bare Vertices}: Example interaction diagrams representing terms in the canonical Yukawa Hamiltonian.}
  \label{fig:bare-vertices}
\end{figure}

The Yukawa Hamiltonian is written as a sum of three terms: 
\begin{equation}
    \label{eq:Ham-terms}
    H = H_0 + H_{\bar\psi \psi\phi} + H_{\bar \psi \phi \phi \psi},
\end{equation}
where $H_0 = H_{\psi^2} + H_{\phi^2}$ is the free kinetic energy of the fermions and bosons respectively, $H_{\bar\psi \psi\phi}$ is the standard Yukawa 3-point interaction, and $H_{\bar \psi \phi \phi \psi}$ is the \textit{instantaneous} interaction term, which is present in light-front coordinates, but not equal time coordinates. Two example diagrams corresponding to $H_{\bar \psi \psi \phi}$ and $H_{\bar \psi \phi \phi \psi}$ are shown in Fig.~\ref{fig:bare-vertices}. 

Standard three-point vertices are marked by a solid black dot, while instantaneous vertices are marked by a short line perpendicular to the ``instantaneous propagator'' in the interaction.

The Hamiltonian can be written in a momentum-space representation.
These Hamiltonian terms are given below~\cite{serafin-yukawa}:

\begin{align}
    \label{eq:H-free}
    H_0 = \frac{m^2}{2}&\int \frac{dq^+}{4\pi} :\bar \psi(q) \frac{\gamma^+}{q^+}\psi(q):\nonumber\\
    &+ \frac{\mu^2}{2}\int \frac{dq^+}{4\pi} :\phi(-q)\phi(q): ,
\end{align}
\begin{align}
    \label{eq:H-3}
    H_{\bar \psi \psi \phi} = g&\int \frac{dq_1^+ dq_2^+ dq_3^+}{(4\pi)^3}4\pi \delta(q_1^+ + q_2^+ +q_3^+)\nonumber \\
    &\times:\bar \psi(-q_1) \psi(q_2) \phi(q_3):,
\end{align}
\begin{align}
    \label{eq:H-4}
    H_{\bar \psi \phi \phi \psi} = g^2&\int \frac{dq_1^+ dq_2^+ dq_3^+ dq_4^+}{(4\pi)^4}4\pi \delta(q_1^+ + q_2^+ +q_3^+ + q_4^+)\nonumber \\
    &\times :\bar \psi(-q_1) \phi(q_2)\frac{\gamma^+/2}{q_3^+ + q_4^+} \phi(q_3) \psi(q_4):,
\end{align}
where $:\hspace{0.3cm}:$ represents \textit{normal ordering}, i.e. ordering all creation operators to the left of annihilation operators. The light-front momentum carried by the $i^{th}$ particle in the interaction is given by $q_i^+$, and $\gamma^+ = \gamma^0 + \gamma^3$, where $\gamma^\mu$ are the Dirac matrices.
Throughout the remainder of this paper, the shorthand
\begin{equation}
    \int_{1,\dots,n} \equiv \int \frac{dq_1^+\dots dq_n^+}{\left( 4\pi\right)^n}
\end{equation}
is used.

The continuous momentum-space fields are:

\begin{equation}
    \label{eq:phi}
    \phi(q) = \frac{\theta(q^+)a_q + \theta(-q^+)a_{-q}^\dagger}{|q^+|},
\end{equation}
\begin{equation}
    \label{eq:psi}
    \psi(q) = \frac{\theta(q^+)u(q)b_q + \theta(-q^+)v(-q)d_{-q}^\dagger}{|q^+|},
\end{equation}
where $\theta(q)$ is the Heaviside function.
The operators $a^\dagger$ and $a$ are bosonic creation and annihilation operators respectively. 
The operators $b^\dagger(d^\dagger)$ and $b(d)$ are fermionic (antifermionic) creation and annihilation operators respectively. 
These operators (referred to collectively as `ladder operators') act on Fock states to create or annihilate particles of a particular type with specific momentum.

In $1+1D$, the representation of the Dirac matrices we use is:
\begin{equation}
    \gamma^0 = \begin{pmatrix}
        0&1\\1&0
    \end{pmatrix},\hspace{1cm} \gamma^3 = \begin{pmatrix}
        0&-1\\1&0
    \end{pmatrix},
\end{equation}
and the spinors, $u$ and $v$ apperaing in Eq.~\ref{eq:psi}, are given by:
\begin{equation}
u(q) = \frac{1}{\sqrt{q^+}}\begin{pmatrix}
    q^+\\m
\end{pmatrix},\hspace{1cm} v(q) = \frac{1}{\sqrt{q^+}}\begin{pmatrix}
    -q^+\\m
\end{pmatrix},
\end{equation}
with conjugate spinors:
\begin{equation}
    \bar u(q) = u^\dagger(q)\gamma^0,\hspace{1cm} \bar v(q) = v^\dagger(q)\gamma^0.
\end{equation}
Relevant spinor identities given our representation of the Dirac matrices are:
\begin{align}
    \bar u(q) u(p) = -\bar v(q) v(p) &= \frac{m}{\sqrt{q^+p^+}}\left(q^+ + p^+\right), \nonumber \\
    \bar u(q) v(p) = -\bar v(q) u(p) &= \frac{m}{\sqrt{q^+p^+}}\left(q^+ - p^+ \right), \nonumber \\
\end{align}
and 
\begin{align}
    \bar u(q) \gamma^+ u(p) = \bar v(q) \gamma^+ v(p) &= {2\sqrt{q^+ p^+}}, \nonumber \\
    \bar u(q)\gamma^+ v(p) = \bar v(q)\gamma^+ u(p) &= -2{\sqrt{q^+p^+}}. \nonumber \\
\end{align}

Inserting~\ref{eq:phi} and~\ref{eq:psi} into Equations~\ref{eq:H-free},~\ref{eq:H-3} and~\ref{eq:H-4} leads to the canonical Yukawa Hamiltonian written as sums of products of creation and annihilation operators. 
The discretization procedure used for this Hamiltonian is Discretized Lightcone Quantization (DLCQ)~\cite{pauli1985discretized,PhysRevD.32.1993}. We will describe this procedure in the next Section~\ref{sec:dlcq}.
\section{Discretized Lightcone Quantization (DLCQ)}
\label{sec:dlcq}

To numerically calculate mass spectra and wavefunctions, a discretization procedure is needed.
The procedure chosen in this work is discretized light-cone quantization (DLCQ)~\cite{PhysRevD.32.1993, pauli1985discretized}.

\subsection{Continuous to discrete theory}
\label{subsec:discrete_hamiltonian}

Following~\cite{PhysRevA.105.032418}, given the total bound state two-momentum $P^\mu$, the invariant momentum squared in $1+1D$ is $P^-P^+ = M^2$.
When this is quantized, so that $P^+$ and $P^-$ become operators, this relation can be interpreted as an eigenvalue equation:
\begin{equation}
    \label{eq:eigenvalue}
    \left(\hat P^- \hat P^+ \right)\ket{\psi} = M^2\ket{\psi}.
\end{equation}
Since $\hat P^+$ and $\hat P^-$ commute, they can be simultaneously diagonalized. 

We take $\ket{\psi}$ to be an eigenstate of $\hat P^+$ with the eigenvalue of the total bound state longitudinal momentum, $P^+$. That is, we diagonalize $\hat P^-$ in a basis of fixed $P^+$.
Eq.~\ref{eq:eigenvalue} then simplifies to 
\begin{equation}
    \label{eq:1p1Deigenvalue}
    \hat P^- \ket{\psi} = \left(\frac{M^2}{P^+}\right)\ket{\psi}.
\end{equation}
While $\hat P^-$ is frame-dependent (it is simply a component of a four-vector), $\hat P^- \hat P^+$ is a Lorentz scalar. 
Consequently, we can choose any value of $P^+>0$ for the total bound state longitudinal momentum of $\ket{\psi}$, diagonalize $\hat P^-$, and expect the same $M^2$ value for any choice of $P^+$.
The light-front energy operator $\hat P^-$ will be denoted as the Hamiltonian, $H$, for the remainder of this paper.


We solve Eq.~\ref{eq:1p1Deigenvalue} by restricting $x^- \in [-L, L]$, i.e. placing the fields in a box. 
The integrals in Eqs.~\ref{eq:H-free},~\ref{eq:H-3}, and~\ref{eq:H-4} over continuous $q_i^+$ become sums over discrete longitudinal momentum, defined as $q^+(k_i)$, where $k_i$ is a momentum quantum number. 
$k$ will be referred to as momentum throughout, even though $q^+(k)$ is the actual momentum and $k$ is the momentum quantum number.

The continuous longitudinal momentum takes on any value from $0$ to $P^+$ in a block of fixed $P^+$, while $q^+(k)$ is discretized such that
\begin{equation}
    q^+(k) = \frac{2\pi}{L}k.
\end{equation}
The momentum quantum number, $k$ in $q^+(k)$ takes on values $\frac12, \frac32, \frac52, \dots$ for (anti-)fermions, and values $1, 2, 3, \dots$ for bosons to satisfy boundary conditions of the fields imposed at $x^- =\pm L$. This follows by the conditions that a bosonic field satisfies periodic boundary conditions, i.e. $\phi(L) = \phi(-L)$, while a fermionic field satisfies antiperiodic boundary conditions i.e. $\psi(L) = -\psi(-L)$.


The continuum limit is recovered by taking $ L \rightarrow  \infty$.
In a bound state calculation, by choosing a finite value of $L$, it necessarily restricts the space of allowed states, due to the positivity of $P^+$, and the fact that the $k_i$'s are discrete. Increasing $L$ leads to a better approximation to the continuum limit.
In practice, $L$ is taken to be integer multiples of $2\pi$, such that $L = 2\pi KL_0$, where $K$ is referred to as the \textit{harmonic resolution}, and $L_0$ is some arbitrary reference length (which in this work is taken to be $L_0 = 1)$. The harmonic resolution can be taken as any value $\frac12, 1, \frac32, \dots$.

The continuum limit ($L \rightarrow \infty$) can now be traded for the limit $K \rightarrow \infty$.
The constraint that all constituent longitudinal momenta $q_i^+$ must sum to the total longitudinal momentum $P^+$ when discretized becomes
\begin{equation}
    P^+ = \frac{1}{K}\sum_i k_i,
\end{equation}
which implies that the bound state constituent momenta quantum numbers $k_i$ in a given Fock state must all sum to $P^+ K$. 

The free fields in momentum space from equations~\ref{eq:phi} and~\ref{eq:psi} are discretized as:
\begin{equation}
    \label{eq:phi-discrete}
    \phi(k) = 2L\frac{\theta(k)a_k + \theta(-k)a_{-k}^\dagger}{\sqrt{4\pi|k|}}
\end{equation}
\begin{equation}
    \label{eq:psi-discrete}
    \psi(k) = 2L\frac{\theta(k)u(k)b_k + \theta(-k)v(-k)d_{-k}^\dagger}{\sqrt{4\pi |k|}}.
\end{equation}
Here, the spinors are written as functions of $k$, the discretized momentum quantum number, where it is assumed implicitly that $u(k) = u(q^+(k))$.
The square root is necessary such that the discretized creation and annihilation operators are normalized to unity: 
\begin{equation}
    \Big[ a_i, a_j^\dagger \Big] = \delta_{ij},
\end{equation}
\begin{align}
   \Bigl\{b_i, b_j^\dagger \Bigr\}=\Bigl\{d_i, d_j^\dagger \Bigr\}= \delta_{ij}.
\end{align}

With the discretization procedure outlined, we now give the discrete form of the canonical Yukawa Hamiltonian.



\subsection{The Discretized Bare Hamiltonian}
\label{subsec:discrete-bare}
As the input to the block encoding described in Section~\ref{be-metrics}, the Hamiltonian must be written as a discrete sum of products of creation and annihilation operators. The free Hamiltonian is written as 

\begin{align}
\label{eq:H-free-discrete}
    H_0 = \frac{L}{2\pi}&\sum_{k \in \mathbb{Z}^+ - 1/2}\left(\frac{m^2}{k} \right)\left(b_{k}^\dagger b_{k} +d_{k}^\dagger d_{k}  \right) \nonumber \\
    +&\frac{L}{2\pi}\sum_{k \in \mathbb{Z}^+} \left(\frac{\mu^2}{k} \right)a_{k}^\dagger a_{k}.
\end{align}
The 3-point standard Yukawa interaction is a sum of six distinct types of terms:
\begin{equation}
    \label{eq:H-3-discrete}
    H_{\bar \psi \psi \phi} = \frac{2Lg}{(4\pi)^{3/2}}\sum_n\sum_{k_1, k_2, k_3}\frac{\delta_{k_1+ k_2 + k_3, 0}}{\sqrt{|k_1k_2k_3|}}:T^{(\mathrm{I})}_n(k_1, k_2, k_3):,
\end{equation}
where $T^{(\mathrm{I})}_n$ corresponds to a term in a row from Table~\ref{table:HY} (i.e. a product of the columns of the $n^{th}$ row).

The column $\theta(k_i, \dots k_n)$ exists in all tables throughout this paper. This notation is shorthand for $\theta(k_i)\dots\theta(k_n)$, where the $\pm$ in a given row of this column corresponds to the sign of $k_i$ inside the Heaviside function.
For example, in Table~\ref{table:HY}, the first row, $-++$ corresponds to $\theta(-k_1)\theta(k_2)\theta(k_3)$.

\begin{table}
  \centering
  \begin{tabular}{||c|c| c| c||} 
 \hline
 $T_n$ & $\theta(k_1,k_2, k_3)$ &Spinors &Operator \\ [0.5ex] 
 \hline\hline
 $T_1$ & $-++$ & $\bar u(-k_1)u(k_2)$ & $b_{-k_1}^\dagger b_{k_2}a_{k_3}$ \\ 
 \hline
 $T_2$ & $-+-$ & $\bar u(-k_1)u(k_2)$ & $b_{-k_1}^\dagger b_{k_2}a_{-k_3}^\dagger$ \\ 
 \hline
 $T_3$ & $+-+$ & $\bar v(k_1)v(-k_2)$ & $d_{k_1} d_{-k_2}^\dagger a_{k_3}$ \\ 
 \hline
 $T_4$ & $+--$ & $\bar v(k_1)v(-k_2)$ & $d_{k_1} d_{-k_2}^\dagger a_{-k_3}^\dagger$ \\ 
 \hline
 $T_5$ & $--+$ & $\bar u(-k_1)v(-k_2)$ & $b_{-k_1}^\dagger d_{-k_2}^\dagger a_{k_3}$ \\ 
 \hline
 $T_6$ & $++-$ & $\bar v(k_1)u(k_2)$ & $d_{k_1}b_{k_2}  a_{-k_3}^\dagger$ \\ 
 \hline
\end{tabular}
  \caption{\justifying \textbf{Discrete} $H_{\bar \psi \psi \phi}$: The standard three-point Yukawa term is a sum of the above 6 rows, where $k_1, k_2 \in \mathbb{Z} + 1/2$, $k_3 \in \mathbb{Z} \backslash \{0\}$, and $\theta(k_1, k_2, k_3) \equiv \theta(k_1)\theta(k_2)\theta(k_3)$.
  The spinors are implicitly assumed to mean $u(k) = u(q^+(k))$.}
  \label{table:HY}
\end{table}

The 4-point instantaneous interaction is 
\begin{align}
\label{eq:H-4-discrete}
H_{\bar \psi \phi \phi \psi} = \frac{2Lg^2}{(4\pi)^2}\sum_n \sum\limits_{\substack{k_1, k_2\\k_3, k_4}}&\frac{\delta_{k_1+ k_2 + k_3+k_4, 0}}{\sqrt{|k_1 k_2 k_3 k_4|}}\nonumber\\
&\times:T^{(\mathrm{II})}_n(k_1, k_2, k_3, k_4):,
\end{align}
where $T^{(\mathrm{II})}_n$ corresponds to a term in a row from Table~\ref{table:HI}.

\begin{table}
  \centering
  \begin{tabular}{||c|c| c| c||} 
 \hline
 $T_n$ & $\theta(k_1,k_2, k_3, k_4)$ &Spinors  &Operator \\ [0.5ex] 
 \hline\hline
 $T_1$ & $-+++$ & $\bar u(-k_1)\gamma^+u(k_4)$ & $b_{-k_1}^\dagger b_{k_4}a_{k_2} a_{k_3}$ \\ 
 \hline
 $T_2$ & $--++$ & $\bar u(-k_1)\gamma^+u(k_4)$ & $b_{-k_1}^\dagger b_{k_4}a_{-k_2}^\dagger a_{k_3}$ \\ 
 \hline
 $T_3$ & $-+-+$ & $\bar u(-k_1)\gamma^+u(k_4)$ & $b_{-k_1}^\dagger b_{k_4}a_{k_2} a_{-k_3}^\dagger$ \\ 
 \hline
 $T_4$ & $---+$ & $\bar u(-k_1)\gamma^+u(k_4)$ & $b_{-k_1}^\dagger b_{k_4}a_{-k_2}^\dagger a_{-k_3}^\dagger$ \\ 
 \hline
 $T_5$ & $+++-$ & $\bar v(k_1)\gamma^+v(-k_4)$ & $d_{k_1} d_{-k_4}^\dagger a_{k_2} a_{k_3}$ \\ 
 \hline
 $T_6$ & $+-+-$ & $\bar v(k_1)\gamma^+v(-k_4)$ & $d_{k_1} d_{-k_4}^\dagger a_{-k_2}^\dagger a_{k_3}$ \\ 
 \hline
 $T_7$ & $++--$ & $\bar v(k_1)\gamma^+v(-k_4)$ & $d_{k_1} d_{-k_4}^\dagger a_{k_2} a_{-k_3}^\dagger$ \\ 
 \hline
 $T_8$ & $+---$ & $\bar v(k_1)\gamma^+v(-k_4)$ & $d_{k_1} d_{-k_4}^\dagger a_{-k_2}^\dagger a_{-k_3}^\dagger$ \\ 
 \hline
 $T_9$ & $ -++-$ & $\bar u(-k_1) \gamma^+ v(-k_4)$ & $b_{-k_1}^\dagger d_{-k_4}^\dagger a_{k_2} a_{k_3}$ \\
 \hline
 $T_{10}$ & $ --+-$ & $\bar u(-k_1) \gamma^+ v(-k_4)$ & $b_{-k_1}^\dagger d_{-k_4}^\dagger a_{-k_2}^\dagger a_{k_3}$ \\
 \hline
 $T_{11}$ & $ -+--$ & $\bar u(-k_1) \gamma^+ v(-k_4)$ & $b_{-k_1}^\dagger d_{-k_4}^\dagger a_{k_2} a_{-k_3}^\dagger$ \\
 \hline
 $T_{12}$ & $+-++$ & $\bar v(k_1) \gamma^+ u(k_4)$ & $ d_{k_1} b_{k_4} a_{-k_2}^\dagger a_{k_3}$\\ \hline
 $T_{13}$ & $++-+$ & $\bar v(k_1) \gamma^+ u(k_4)$ & $ d_{k_1} b_{k_4} a_{k_2} a_{-k_3}^\dagger$\\ \hline
 $T_{14}$ & $+--+$ & $\bar v(k_1) \gamma^+ u(k_4)$ & $ d_{k_1} b_{k_4} a_{-k_2}^\dagger a_{-k_3}^\dagger$\\ \hline
\end{tabular}
  \caption{\justifying \textbf{Discrete} $H_{\bar \psi \phi \phi \psi}$: The instantaneous Yukawa Hamiltonian is a sum of the terms in the above 14 rows, where $k_1, k_4 \in \mathbb{Z} + 1/2$, $k_2, k_3 \in \mathbb{Z} \backslash \{0\}$, and $\theta(k_1, k_2, k_3, k_4) \equiv \theta(k_1)\theta(k_2)\theta(k_3)\theta(k_4)$. The spinors are implicitly assumed to mean $u(k) = u(q^+(k))$.
  }
  \label{table:HI}
\end{table}

After fixing $P^+$, one diagonalizes the Hamiltonian at a particular $K$ to obtain mass eigenvalues $M^2$. Increasing $K \to\infty$ leads to a nonphysical and/or divergent spectrum, see Fig.~\ref{fig:bare_spectrum}. 
This is because at higher harmonic resolutions, effects from higher order diagrams, e.g., loops,  begin to dominate.
A systematic method to handle and remove these divergences is needed in order to obtain physical results. 
In this paper, this is achieved via the Renormalization Group Procedure for Effective Particles (RGPEP)~\cite{glazek2012perturbativeformulaerelativisticinteractions, GLAZEK2000175}.

For the remainder of this paper, $P^+$ is taken to be $1$ without loss of generality.
\section{Renormalization Group Procedure for Effective Particles (RGPEP)}
\label{sec:rgpep}
In this section, we describe the Renormalization Group Procedure for Effective Particles (RGPEP). For a thorough introduction, see Ref.~\cite{serafin-yukawa}.

\subsection{Overview}
\label{sec:rgpep-overview}
Divergences appearing in quantum field theories in a Hamiltonian approach can arise from interactions between particles of vastly different energy scales~\cite{Wilson:1965zzb}. After choosing a basis of Fock states, interactions between such distant energy scales manifest as far off-diagonal Hamiltonian matrix elements. 

The similarity renormalization group~\cite{PhysRevD.48.5863, szpigel2000similarity} aims to fix this by applying a unitary transformation $U(s)$ to a matrix $H$, resulting in a new matrix $H(s)$: 

\begin{equation}
\label{eq:srg}
    H(s) = U^\dagger(s)HU(s),
\end{equation}
where $s \in [0, \infty)$ continuously parameterizes the transformation, starting from $s = 0$ such that $H(s = 0) = H$, i.e., $U(0) = U^\dagger(0) = \mathds{1}$.
 Notably, $H(s)$ shares the same spectrum as $H$ as a consequence of the similarity transformation.

 The evolution of $H(s)$ with increasing $s$ comes from differentiating both sides of Eq.~\ref{eq:srg}, leading to:
 \begin{equation}
     \label{eq:rgpep-ut}
    \frac{dH(s)}{ds} = \left[\frac{dU^\dagger(s)}{ds}U(s), H(s) \right].
 \end{equation}
 The first term in the commutator, is called the \textit{generator},  $\mathcal{G}(s)$ and is an anti-Hermitian operator.

In general, the choice of generator is arbitrary, in the same way that $U(s)$ is.
In this work, we use the generator:
\begin{equation}
\label{eq:rgpep-generator}
    \mathcal{G}(s) \equiv \Big[H_0, H(s) \Big],
\end{equation}
where $H_0$ is the free ($g = 0$) part of the Hamiltonian $H(s)$.
This generator is referred to as the RGPEP generator.

With this choice of generator, the Renormalization Group Procedure for Effective Particles (RGPEP) equation is given as 

\begin{equation}
    \label{eq:rgpep}
    \frac{dH(s)}{ds} = \Bigg[\Big[H_0, H(s) \Big], H(s) \Bigg].
\end{equation}

In quantum field theories, Hamiltonians are written as integrals of fields, built out of ladder operators. 
Given a set of Fock states, one could construct a matrix, and solve Eq.~\ref{eq:rgpep}; however, RGPEP offers an alternative approach.

In RGPEP, rather than calculating commutators of matrices, one calculates commutators of creation and annihilation operators. 
In general, the RGPEP equation cannot be solved exactly, so it must be solved order-by-order in $g$ (see~\cite{exactlysolvable} for an exactly solvable model).

The RGPEP equation is solved perturbatively by expanding the \textit{effective} Hamiltonian $H(s)$ in powers of the coupling:
\begin{equation}
    \label{eq:H-expansion}
    H(s) = H^{(0)}(s) + gH^{(1)}(s) + g^2 H^{(2)}(s) + \mathcal{O}(g^3).
\end{equation} 
Even though $\mathcal{O}(g^3)$ terms do not exist in the canonical Hamiltonian, $H(s)$ includes higher order terms. 
This is because upon diagonalization of the bare Hamiltonian, higher order diagrams arise, and must be explicitly accounted for in the effective Hamiltonian, $H(s)$.

Asserting this form of $H(s)$ in Eq.~\ref{eq:rgpep}, we obtain a differential equation at each order in the coupling:
\begin{subequations}   

    \begin{align}
        &\mathcal{O}(g^0): \frac{dH^{(0)}(s)}{ds} = 0,  \label{eq:rgpep-order-by-order-a} \\
        &\mathcal{O}(g^1): g\frac{dH^{(1)}(s)}{ds} = \Bigg[\Big[H_0(s), gH^{(1)}(s)\Big], H_0(s)\Bigg],\label{eq:rgpep-order-by-order-b}\\
        &\mathcal{O}(g^2): g^2\frac{dH^{(2)}(s)}{ds} = \Bigg[\Big[H_0(s), g^2H^{(2)}(s)\Big], H_0(s)\Bigg] \nonumber \\
        &\hspace{1cm} + \Bigg[\Big[H_0(s), gH^{(1)}(s)\Big],gH^{(1)}(s)\Bigg], \label{eq:rgpep-order-by-order-c} \\
        &\hspace{3cm}\vdots\nonumber
    \end{align}
    \end{subequations}
with initial condition 
\begin{equation}
    H(s = 0) = H.
\end{equation}
The solutions of each of these equations give the $n^{th}$ order terms in the effective Hamiltonian. 
The $n^{th}$ order RGPEP equation depends on solutions to all orders $\leq n$. 
To solve these differential equations, we follow the approach of~\cite{serafin-yukawa}, which is summarized in Appendix~\ref{sec:orders}.

The parameter $s$ can be replaced with $\lambda$ by the change of variables $\lambda = 1/\sqrt{s}$, such that the units are consistent.
This change of variables is useful because $\lambda$ has a nice physical interpretation:
$\lambda$ is interpretted as an energy scale, such that decreasing $\lambda$ ``lowers the energy scale" at which interactions may occur (by mitigating far off-diagonal matrix elements).

\begin{figure}
    \centering
    \includegraphics[width=0.75\linewidth]{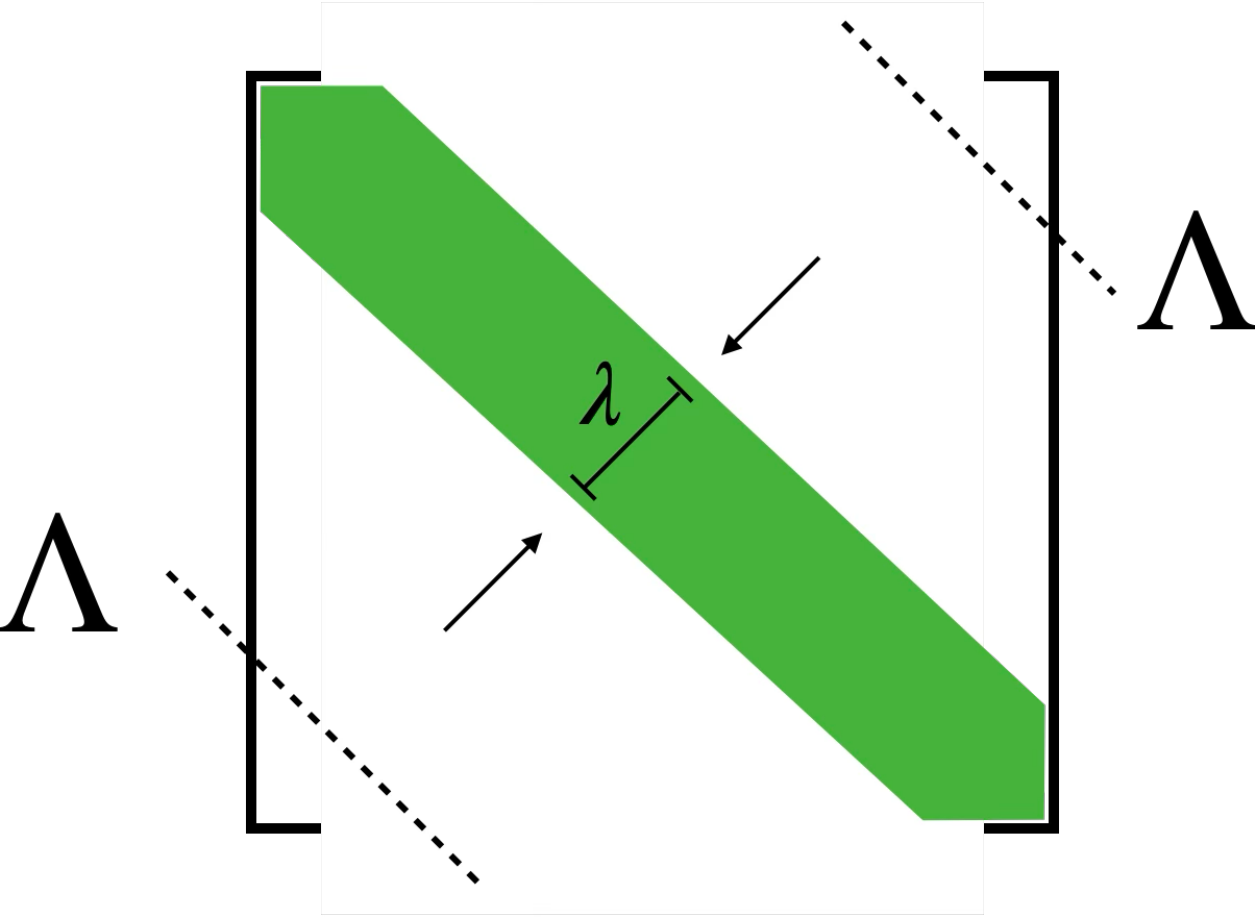}
    \caption{\justifying\textbf{Effective Hamiltonian} $H(\lambda)$, regulated by $\Lambda$, which acts as a cutoff for far off-diagonal matrix elements. $\lambda$ is the renormalization scale and determines the width of the diagonal band $U(\lambda)$ produces. This figure shows the Hamiltonian flow with decreasing $\lambda$.}
    \label{fig:H(lambda)}
\end{figure}

A regulator, $\Lambda$, which mitigates far off-diagonal matrix elements, is imposed through a regulating function. 
The regulator $\Lambda$ makes the theory finite and well-defined, but the results depend strongly on the choice of the regulator. One aims to renormalize the theory by making it insensitive to this choice, i.e. allowing one to remove the regulator by taking the limit $\Lambda \rightarrow \infty$, while observables remain finite.
 The evolution of the regulated bare Hamiltonian under the flow is shown in Fig.~\ref{fig:H(lambda)}.
Colloquially, Fig.~\ref{fig:H(lambda)} shows the Hamiltonian `flowing' towards the diagonal with decreasing $\lambda$.

For the remainder of the paper, the renormalization scale parameter is taken to be $\lambda$, with all Hamiltonians given with $\lambda$-dependence.

\subsection{The Effective Yukawa Hamiltonian}
Here, the solutions to the differential equations~\ref{eq:rgpep-order-by-order-a},~\ref{eq:rgpep-order-by-order-b} and~\ref{eq:rgpep-order-by-order-c} are given.
\subsubsection{$\mathcal{O}\left(g^0\right)$ Solution}
At $\mathcal{O}\left(g^0\right)$, the solution to the RGPEP equation is:
\begin{equation}
    \label{eq:zeroth_order}
    H^{(0)}(\lambda) = H_0,
\end{equation}
showing that the free Hamiltonian is unaffected by the RGPEP unitary evolution.
\subsubsection{$\mathcal{O}\left(g^1\right)$ Solution}
The $\mathcal{O}\left(g\right)$ solution of the RGPEP equation is the same as the first order interaction term in the canonical Hamiltonian, Eq.~\ref{eq:H-3}, modified by a \textit{form factor}: 
\begin{equation}
    \label{eq:form-factor}
    f(\lambda; \mathcal{Q}^-) = e^{-\left(\mathcal{Q}^-/\lambda\right)^2}
\end{equation}
such that
\begin{align}
    \label{eq:first_order}
    H^{(1)}(\lambda) &=  \int\limits_{1,2,3}4\pi\delta(\mathcal{Q}^+)f(\lambda;\mathcal{Q}^-)r(\Lambda;\mathcal{Q}^-)\nonumber \\
    &\hspace{2cm}\times:\bar \psi(-q_1) \psi(q_2) \phi(q_3):\nonumber\\
    &\equiv H_{\bar \psi \psi \phi}(\lambda).
\end{align}
For a 3-point vertex, 
\begin{equation}
    f(\lambda; \mathcal{Q}^-)= e^{-\left(q_1^- + q_2^- + q_3^-\right)^2/\lambda^2}
\end{equation}
is the explicit form of the form factor.

\begin{figure}
    \centering
    \includegraphics[width=0.61\linewidth]{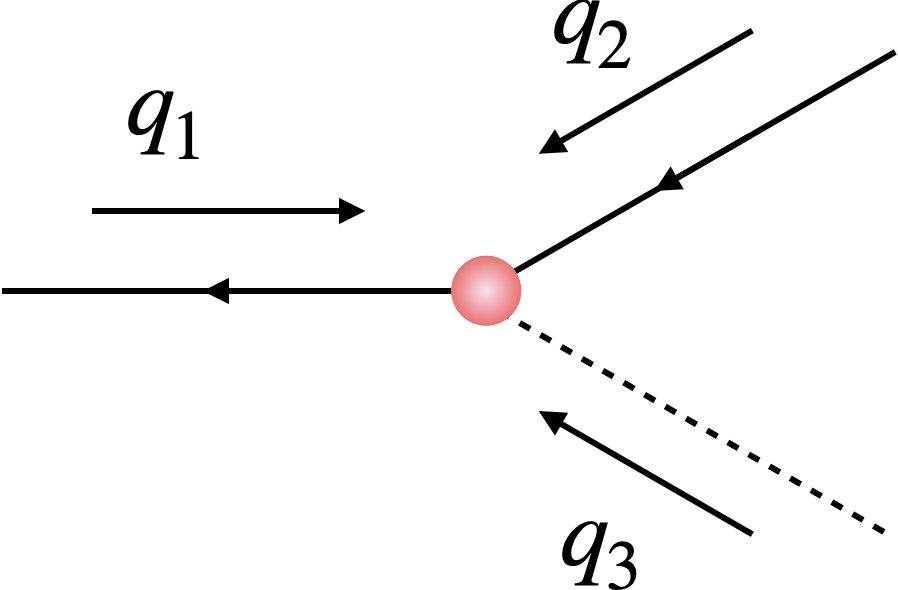}
    \caption{\justifying\textbf{Effective first-order vertex.} An example first order \textit{effective} vertex contributing to $H^{(1)}(\lambda)$.}
    \label{fig:first_order}
\end{figure}

We represent first order \textit{effective} vertices (those modified by a form factor) as the initial vertices with larger red dots, see Fig.~\ref{fig:first_order}.

In Eq.~\ref{eq:first_order}, $r(\Lambda; \mathcal{Q}^-)$ is the regulating function, which regulates $H$ by exponentially dampening far off-diagonal matrix elements. 
Its form is taken to be identical to $f(\lambda; \mathcal{Q}^-)$ but with $\lambda$ replaced by $\Lambda$:
\begin{equation}
    r(\Lambda; \mathcal{Q}^-) = e^{-(q_1^- + q_2^- + q_3^-)^2/\Lambda^2}.
\end{equation}
This is a useful regulating function because it can be combined easily with the form factor.

\subsubsection{$\mathcal{O}\left(g^2\right)$ Solution}
\begin{figure}
    \centering
    \includegraphics[width=0.61\linewidth]{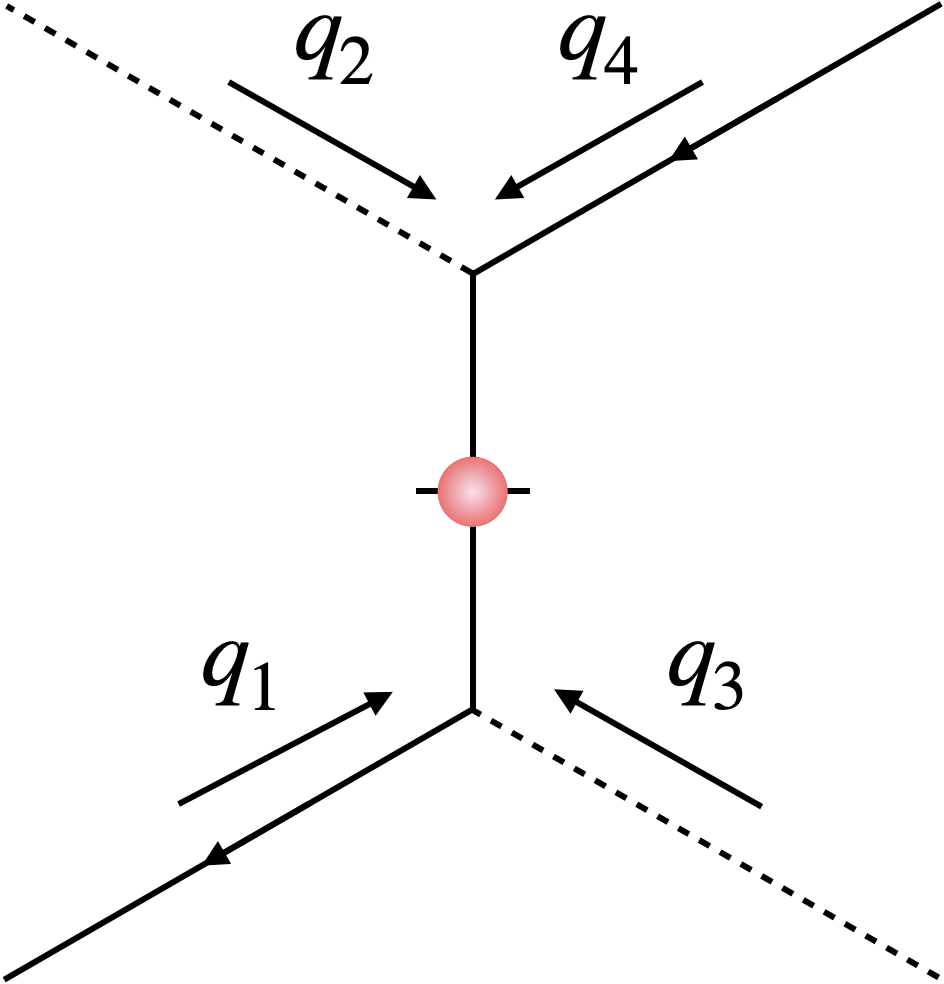}
    \caption{\justifying\textbf{Effective second-order vertex} An example second order \textit{effective} vertex contributing to $H^{(2)}(\lambda)$.}
    \label{fig:effective_4pt}
\end{figure}

At $\mathcal{O}\left(g^2 \right)$, the effective Hamiltonian has contributions from terms in the canonical Hamiltonian and new terms. The second order effective Hamiltonian can be written as: 
\begin{align}
\label{eq:second_order}
    H^{(2)}(\lambda) &= H_{\bar \psi \phi \phi \psi}\left(\lambda\right) + H_{\text{fe}}(\lambda) + H_{\text{be}}(\lambda)  \nonumber\\
    &+ H_{\delta{m^2}}(\lambda) + H_{\delta{\mu^2}}(\lambda) + X_{\delta{m^2}}  + X_{\delta{\mu^2}}.
\end{align}
The first term in Eq.~\ref{eq:second_order} is analogous to 
Eq.~\ref{eq:first_order} in that the $H_{\bar \psi \phi \phi \psi}$ terms present in the canonical Hamiltonian are modified by a form factor, which can be seen in Fig.~\ref{fig:effective_4pt}.
In our compact notation:
\begin{align}
    \label{eq:HI(lambda)}
    H_{\bar \psi \phi \phi \psi}(\lambda) = &\int\limits_{1,2,3,4}4\pi\delta(\mathcal{Q}^+)f(\lambda;\mathcal{Q}^-)r(\Lambda;\mathcal{Q}^-)  \nonumber\\
    &\times :\bar \psi(-q_1) \phi(q_2)\frac{\gamma^+/2}{q_3^+ + q_4^+} \phi(q_3) \psi(q_4):,
\end{align}
where for this interaction vertex, $\mathcal{Q}^\pm = q_1^\pm + q_2^\pm + q_3^\pm +q_4^\pm$.

\begin{figure}
    \centering
    \includegraphics[width=\linewidth]{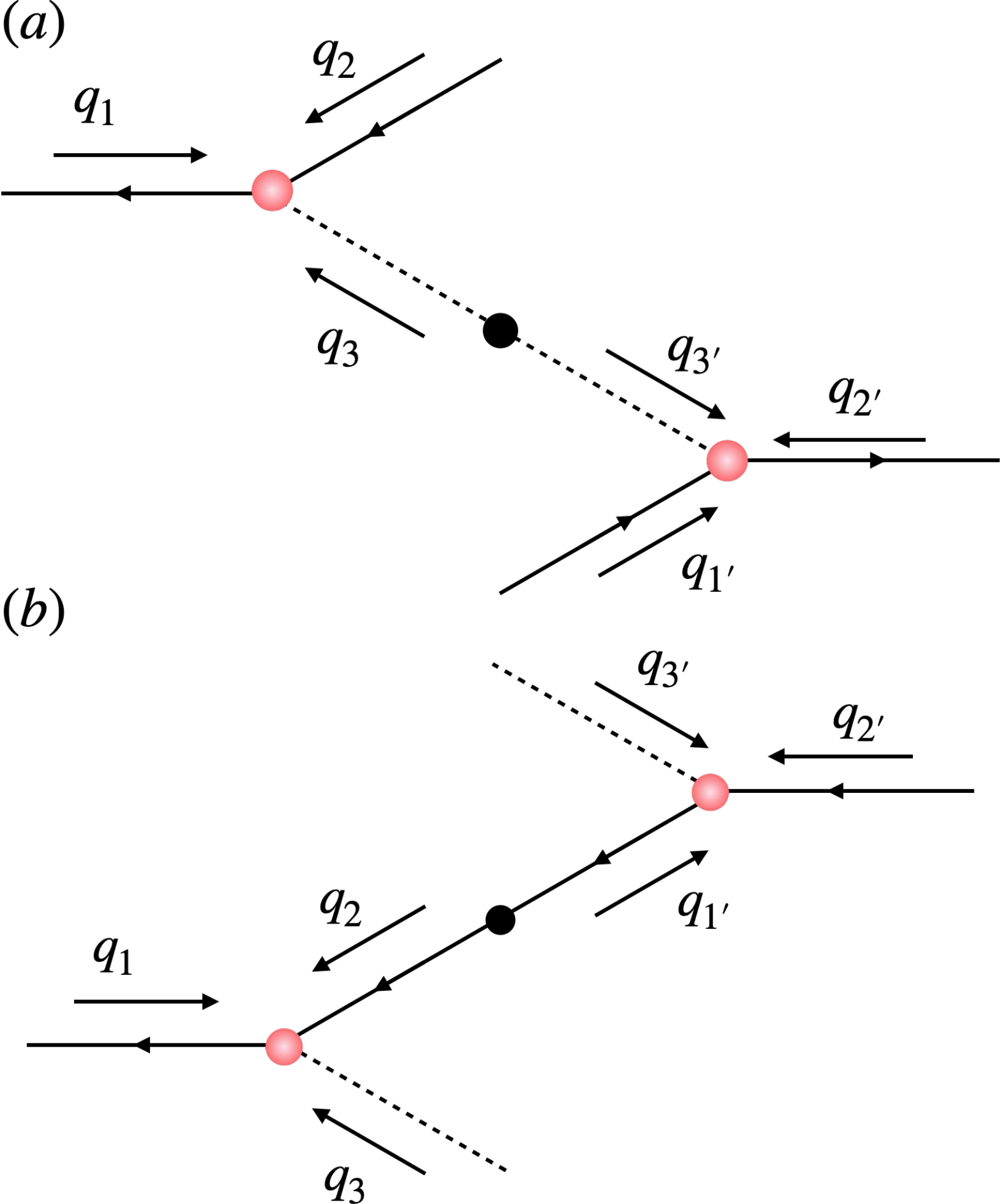}
    \caption{\justifying\textbf{Contracted Terms} Two example exchange terms arising from solving the RGPEP equation at $\mathcal{O}\left(g^2\right)$. (a) An example contraction of two external boson legs contributing to $H_{\text{be}}(\lambda)$. Here, $\mathcal{Q}^- = q_1^- + q_2^- + q_3^-$ and $\mathcal{Q'}^- = q_{1'}^- + q_{2'}^- + q_{3'}^-$. The contraction leads to a Dirac delta $\delta \left(q_3^+ + q_{3'}^+ \right)$, which, under the $q_{3'}$ integration, $\mathcal{Q'}$ can be written as $\mathcal{Q'}^- = q_{1'}^- + q_{2'}^- - q_{3}^-$ since $q_3^- = \mu^2/q_3^+$. (b) An example contraction of two external fermion legs contributing to $H_{\text{fe}}(\lambda)$. Here, $\mathcal{Q}^- = q_1^- + q_2^- + q_3^-$ and $\mathcal{Q'}^- = q_{1'}^- + q_{2'}^- + q_{3'}^-$. The contraction leads to a Dirac delta $\delta \left(q_2^+ + q_{1'}^+ \right)$, which, under the $q_{1'}$ integration, $\mathcal{Q'}^-$ can be written as $\mathcal{Q'}^- = q_{2'}^- + q_{3'}^- - q_2^-$ since $q_2^- = m^2/q_2^+$}
    \label{fig:contractions}
\end{figure}

\begin{figure}
    \centering
    \includegraphics[width=0.75\linewidth]{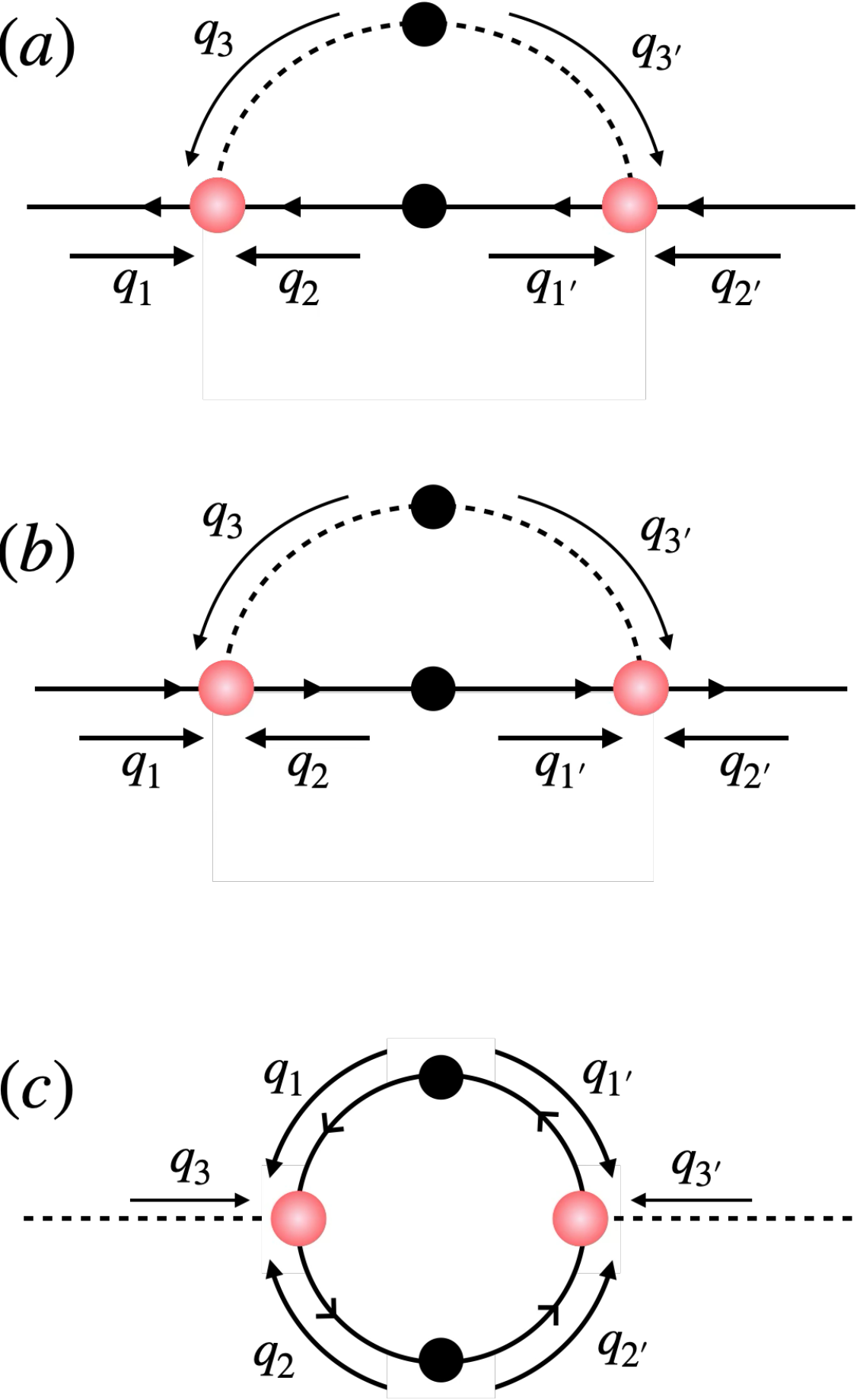}
    \caption{\justifying\textbf{Loop Diagrams} double contraction terms arising from solving the RGPEP equation at $\mathcal{O}\left(g^2\right).$ (a)  A double contraction between external boson and fermion legs leading to a fermion mass loop diagram. (b) A double contraction between external boson and antifermion legs leading to an antifermion mass loop diagram. (c) A double contraction between external fermion and antifermion legs leading to a boson mass loop diagram.}
    \label{fig:loops}
\end{figure}

The second commutator in Eq.~\ref{eq:rgpep-order-by-order-c} contains $H^{(1)}(\lambda)$ twice, and can be evaluated using Wick's theorem~\cite{serafin-yukawa}.
Wick's theorem requires calculation of all possible \textit{contractions}. A contraction is defined as:
\begin{equation}
    \wick{\c A \c B} = AB\hspace{1mm} - :AB:.
\end{equation}
The non-zero contractions of $:\bar \psi \psi \phi:$ with $:\bar \psi \psi \phi:$ (the product of two first order interactions)  are:
\begin{align}
    \label{eq: contractions}
    &\wick[offset=1.5em]{:\c{\bar \psi} \psi \phi \bar \psi \c\psi \phi:}, \wick[offset=1.5em]{:{\bar \psi} \c{\psi}\phi \c{\bar \psi} \c\psi \phi:}, \wick[offset=1.5em]{:{\bar \psi} \psi \c\phi \bar \psi \psi \c\phi:},\nonumber \\
    &\wick[offset=1.5em]{:{\bar \psi} \c1\psi \c2\phi \c1{\bar \psi }\psi \c2\phi:}, 
    \wick[offset=1.5em]{:\c1{\bar \psi} \psi \c2\phi {\bar \psi }\c1\psi \c2\phi:},\wick[offset=1.5em]{:\c1{\bar \psi} \c2\psi \phi \c2{\bar \psi }\c1\psi \phi:},
\end{align}
where the single contractions represent the exchange terms, $H_{\text{be}}(\lambda)$ and $H_{\text{fe}}(\lambda)$, depicted in Fig.~\ref{fig:contractions}, while the double contractions (two contractions in a single product) represent the loop diagrams, $H_{\delta m^2}(\lambda)$ and $H_{\delta \mu^2}(\lambda)$, shown in Fig.~\ref{fig:loops}. 

Lastly, $X_{\delta{m^2}}$and $X_{\delta{\mu^2}}$ are counterterms added to cancel divergences arising from the loop diagrams. The forms of the counterterms are given in Appendix~\ref{sec:counterterms}.

Before giving the explicit forms of the exchange and loop diagrams, it is useful to picture a contraction as stitching two external legs in a diagram together. Therefore, one takes two first-order interaction vertices in the canonical regulated Hamiltonian and labels one with primes to differentiate the two, as shown in Fig.~\ref{fig:contractions}. 

The fermion exchange term is given as:
\begin{align}
    H_{\text{fe}}(\lambda) &=  \int\limits_{\substack{1,2,3\\1'2'3'}}4\pi \delta(\mathcal{Q}^+)4\pi \delta(\mathcal{Q'}^+)4\pi\delta(q_2^++q_{1'}^+)\nonumber \\
&\times\frac{1}{q_2^+}\mathcal{R}\Big(\lambda;\mathcal{Q}^-, \mathcal{Q}'^-\Big)\nonumber \\
    &\times:\bar \psi(-q_1)\phi(q_3)\left(\gamma_{\mu}q_2^\mu + m\right)\phi(q_{3'})\psi(q_{2'}):,
\end{align}
while the bosonic exchange term is:
\begin{align}
    H_{\text{be}}(\lambda) = \int\limits_{\substack{1,2,3\\1'2'3'}}&4\pi \delta(\mathcal{Q}^+)4\pi \delta(\mathcal{Q'}^+)4\pi\delta(q_3^++q_{3'}^+)\nonumber \\
&\times\frac{\theta(q_3^+)}{q_3^+}\mathcal{R}\Big(\lambda;\mathcal{Q}^-, \mathcal{Q}'^-\Big)\nonumber \\
    &\times:\bar \psi(-q_1)\psi(q_2) \bar \psi(-q_1')\psi(q_2'):.
\end{align}
$\mathcal{R}$ is a function of the relevant momenta in the interaction that comes from the RGPEP solutions. It is given by:
\begin{align}
    \mathcal{R}&\Big(\lambda; \mathcal{Q}^-, \mathcal{Q}'^-\Big) = \frac12 \Big(\frac{1}{\mathcal{Q}^-} - \frac{1}{\mathcal{Q}'^-}\Big)\nonumber \\
    &\times\Big( f(\lambda;\mathcal{Q}^-)\cdot f(\lambda;\mathcal{Q}'^-) - f(\lambda;\mathcal{Q}^- + \mathcal{Q}'^-)\Big)\nonumber \\
    &\times r(\Lambda;\mathcal{Q}^-)r(\Lambda; \mathcal{Q}'^-).
\end{align}


Double contractions lead to loop diagrams:
\begin{align}
    H_{\delta m^2}&(\lambda) =  \frac12\int\limits_{q}\delta m^2:\bar \psi(q)\frac{\gamma^+}{q^+}\psi(q):,
\end{align}
where 
\begin{align}
\label{eq:fermion-loop}
    \delta m^2 &= \int_0^1 \frac{dx}{x(1-x)}\Big(\mathcal{M}_{fb}^2(x) -  m^2\Big)^{-1}\nonumber \\
    &\times \frac{m^2(1+x)^2}{x}\Bigg(e^{-2\Big(\mathcal{M}_{fb}^2(x) - m^2\Big) ^2/(q^+\lambda)^2}  - 1\Bigg)\nonumber \\
    &\times e^{-2\Big(\mathcal{M}_{fb}^2(x) - m^2\Big) ^2/(q^+\Lambda)^2}
\end{align}
and 
\begin{align}
    H_{\delta  \mu^2}&(\lambda) =\frac12 \int\limits_{q}\delta \mu^2 :\phi(-q)\phi(q):,
\end{align}
where
\begin{align}
\label{eq:boson-loop}
    \delta \mu^2 &= \int_0^1 \frac{dx}{x(1-x)}\Big(\mathcal{M}_{f\bar f}^2(x) -  \mu^2\Big)^{-1}\nonumber \\
    &\times \frac{m^2(2x-1)^2}{x(1-x)}\Bigg(e^{-2\Big(\mathcal{M}_{f \bar f}^2(x) - \mu^2\Big) ^2/(q^+\lambda)^2}  - 1\Bigg)\nonumber \\
    &\times e^{-2\Big(\mathcal{M}_{f\bar f}^2(x) - \mu^2\Big) ^2/(q^+\Lambda)^2}.
\end{align}
The loops can be seen in Fig.~\ref{fig:loops}.

The invariant mass of two particles, $i$ and $j$, $\mathcal{M}_{ij}^2(x)$ is:
\begin{equation}
\label{eq:inv-mass}
    \mathcal{M}_{ij}^2(x) = \frac{m_i^2}{x} + \frac{m_j^2}{1-x},
\end{equation}
where $m_f = m_{\bar f} = m$ and $m_b = \mu$.
The loop integrals $\delta m^2$, and $\delta \mu^2$ are finite for finite $\Lambda$, but diverge in the limit $\Lambda \rightarrow \infty$.

A step-by-step method of solving equations~\ref{eq:rgpep-order-by-order-a},~\ref{eq:rgpep-order-by-order-b} and~\ref{eq:rgpep-order-by-order-c} can be found in~\cite{serafin-yukawa} and Appendix~\ref{sec:orders}.
Additionally, the form of the counterterms, which cancel the divergences in the $\delta m^2$ and $\delta \mu^2$ integrals, are given in appendix~\ref{sec:counterterms}. One aims to choose counterterms such that observables become finite, and the regulator in $H(\lambda)$ can be removed, i.e. $\Lambda \rightarrow \infty$.
\subsection{Fixing Counterterms}
The final step in any renormalization procedure is to fix the finite part of the counterterms. The general approach to fix the counterterms is to use experimental data to match some observable in the theory. From here, one can make non-trivial predictions of other observables with a finite well-defined theory.

In QCD, fixing counterterms amounts to fitting the parameters of the theory to experimental hadronic measurements. For Yukawa theory, rather than fiting parameters to experimental data, a renormalization condition is set that says the effective single fermion/boson mass must be equal to the corresponding parameter in the Lagrangian. A more thorough discussion on this choice of renormalization condition takes place in Appendix~\ref{sec:counterterms} and Ref.~\cite{serafin-yukawa}.

\subsection{The Discretized Effective Yukawa Hamiltonian}
The discretized form of the renormalized Hamiltonian (without the regulator) can be written in the same way that the discrete bare Hamiltonian is written in Section~\ref{subsec:discrete-bare}. $H_0$ in the effective Hamiltonian is the same as $H_0$ in the bare Hamiltonian, so its discretized form is given in 
Eq.~\ref{eq:H-free-discrete}. 

The discrete `energy' is:
\begin{equation}
    q^-(k_i) = \frac{m_i^2}{q^+(k_i)} = \Big(\frac{L}{2\pi} \Big)\frac{m_i^2}{k_i} = \frac{L}{2\pi}k_i^-.
\end{equation}
From this, the discretized form of $H^{(1)}(\lambda)$ is the same as the discretized 3-point interaction in Eq.~\ref{eq:H-3-discrete}, but modified by the discretized form factor:
\begin{align}
    \label{eq:effective-H-3-discrete}
    H^{(1)}(\lambda) &= \frac{2L}{(4\pi)^{3/2}}\sum_n\sum_{k_1, k_2, k_3}\frac{\delta_{k_1 + k_2 + k_3, 0}}{\sqrt{|k_1k_2k_3|}}\nonumber\\
    &\times e^{-L^2(k_1^- + k_2^- + k_3^- )^2/4\pi^2\lambda^2}:T^{(\mathrm{I})}_n(k_1, k_2, k_3):,
\end{align}
where $k_1$ and $k_2$ are the fermion momenta and $k_3$ is the boson momentum.

 Similarly, the discretized form of $H_{\bar \psi \phi \phi \psi}$ is the same as Eq.~\ref{eq:H-4-discrete}, but again, modified by the discretized form factor: 
 \begin{align}
\label{eq:effective-H-4-discrete}
H_{\bar \psi \phi \phi \psi}&(\lambda) = \frac{2L}{(4\pi)^2}\sum_n \sum\limits_{\substack{k_1, k_2\\k_3, k_4}}\frac{\delta_{k_1+ k_2 + k_3+k_4, 0}}{\sqrt{|k_1 k_2 k_3 k_4|}}\nonumber\\
&\times e^{-L^2(k_1^- + k_2^- + k_3^-+k_4^-)^2/4\pi^2\lambda^2}:T^{(\mathrm{II})}_n(k_1, k_2, k_3, k_4):,
\end{align}
where $k_1$ and $k_4$ are the fermion momenta while $k_2$ and $k_3$ are the boson momenta.
It is important to note that Eq.~\ref{eq:effective-H-3-discrete} makes up the entirety of $H^{(1)}(\lambda)$, while Eq.~\ref{eq:effective-H-4-discrete} is only a part of $H^{(2)}(\lambda)$.

The discretized boson exchange term, $H_{\text{be}}(\lambda)$, is given as 
\begin{align}
\label{eq:Hbe-discrete}
    H_{\text{be}}(\lambda) &= \frac{1}{2L(4\pi)^4}\sum_n \sum\limits_{\substack{k_1, k_2, k_3\\k_{1'}, k_{2'}}}\frac{\theta(k_3)\delta_{k_1, -k_2-k_3}\delta_{k_3, k_{1'} + k_{2'}}}{\sqrt{|k_1k_2k_{1'}k_{2'}|}}\nonumber \\
    &\times \mathcal{R}\Bigg(\lambda;\Big(q^-(k_1) + q^-(k_2) + q^-(k_3)\Big), \nonumber\\
    &\hspace{2cm}\Big(q^-(k_{1'}) +  q^-(k_{2'}) - q^-(k_3)\Big)\Bigg)\nonumber\\&\times:T^{(\mathrm{III})}_n(k_1, k_2, k_{1'}, k_{2'}):,
\end{align}
where $k_1, k_2, k_{1'}, k_{2'}$ correspond to the fermion momenta while $k_3$ corresponds to the exchanged boson momentum. $T^{(\mathrm{III})}_n$ corresponds to a term in a row in Table~\ref{table:Hbe}.

\begin{table}
  \centering
  \begin{tabular}{||c|c| c| c||} 
 \hline
 $T_n$ & $\theta$ &Spinors &Operator \\ [0.5ex] 
 \hline\hline
 $T_1$ & $-+-+$ & $\bar u(-k_1)u(k_2)\bar u(-k_{1'})u(k_{2'})$ & $b_{-k_1}^\dagger b_{k_2} b^\dagger_{-k_{1'}} b_{k_{2'}}$\\ \hline
 $T_2$ & $-+--$ & $\bar u(-k_1)u(k_2)\bar u(-k_{1'})v(-k_{2'})$ & $b_{-k_1}^\dagger b_{k_2} b^\dagger_{-k_{1'}} d_{-k_{2'}}^\dagger$\\ \hline
 $T_3$ & $-+++$ & $\bar u(-k_1)u(k_2)\bar v(k_{1'})u(k_{2'})$ & $b_{-k_1}^\dagger b_{k_2} d_{k_{1'}} b_{k_{2'}}$\\ \hline
 $T_4$ & $-++-$ & $\bar u(-k_1)u(k_2)\bar v(k_{1'})v(-k_{2'})$ & $b_{-k_1}^\dagger b_{k_2} d_{k_{1'}} d_{-k_{2'}}^\dagger$\\ \hline
 $T_5$ & $+--+$ & $\bar v(k_1) v(-k_2) \bar u(-k_{1'}) u(k_{2'})$ & $d_{k_1}d_{-k_2}^\dagger b_{-k_{1'}}^\dagger b_{k_{2'}}$\\ \hline
  $T_6$ & $+---$ & $\bar v(k_1) v(-k_2) \bar u(-k_{1'}) v(-k_{2'})$ & $d_{k_1}d_{-k_2}^\dagger b_{-k_{1'}}^\dagger d_{-k_{2'}}^\dagger$\\ \hline
  $T_7$ & $+-++$ & $\bar v(k_1) v(-k_2) \bar v(k_{1'}) u(k_{2'})$ & $d_{k_1}d_{-k_2}^\dagger d_{k_{1'}} b_{k_{2'}}$\\ \hline
  $T_8$ & $+-+-$ & $\bar v(k_1) v(-k_2) \bar v(k_{1'}) v(-k_{2'})$ & $d_{k_1}d_{-k_2}^\dagger d_{k_{1'}} d_{-k_{2'}}^\dagger$\\ \hline
  $T_9$ & $---+$ & $\bar u(-k_1) v(-k_2) \bar u(-k_{1'}) u(k_{2'})$ & $b_{-k_1}^\dagger d_{-k_2}^\dagger b_{-k_{1'}}^\dagger b_{k_{2'}}$\\ \hline
  $T_{10}$ & $--++$ & $\bar u(-k_1) v(-k_2) \bar v(k_{1'}) u(k_{2'})$ & $b_{-k_1}^\dagger d_{-k_2}^\dagger d_{k_{1'}} b_{k_{2'}}$\\ \hline
  $T_{11}$ & $--+-$ & $\bar u(-k_1) v(-k_2) \bar v(k_{1'}) v(-k_{2'})$ & $b_{-k_1}^\dagger d_{-k_2}^\dagger d_{k_{1'}} d_{-k_{2'}}^\dagger$\\ \hline
  $T_{12}$ &$++-+$ & $\bar v(k_1) u(k_2)\bar u(-k_{1'})u(k_{2'}) $ & $d_{k_1} b_{k_2} b_{-k_{1'}}^\dagger b_{k_{2'}}$ \\ \hline
  $T_{13}$ &$+++-$ & $\bar v(k_1) u(k_2)\bar v(k_{1'})v(-k_{2'}) $ & $d_{k_1} b_{k_2} d_{k_{1'}} d_{-k_{2'}}^\dagger$ \\ \hline
  $T_{14}$ &$++--$ & $\bar v(k_1) u(k_2)\bar u(-k_{1'})v(-k_{2'}) $ & $d_{k_1} b_{k_2} b_{-k_{1'}}^\dagger d_{-k_{2'}}^\dagger$ \\ \hline
  
\end{tabular}
  \caption{\justifying \textbf{Discrete} $H_\text{be}(\lambda)$: $k_i \in \mathbb{Z} + 1/2$ $\forall i$. The column $\theta$ is simplified as $\theta \equiv \theta(k_1,k_2, k_{1'}, k_{2'})$ for brevity. The spinors are implicitly assumed to mean $u(k) = u(q^+(k))$.
  }
  \label{table:Hbe}
\end{table}

\begin{table}
  \centering
  \begin{tabular}{||c|c| c| c||} 
 \hline
 $T_n$ & $\theta$ &Spinors &Operator \\ [0.5ex] 
 \hline\hline
 $T_1$ & $-+++$ &$\bar u(-k_1)(\gamma_\mu {q_2}^\mu + m) u(k_{2'})$ & $b_{-k_1}^\dagger b_{k_{2'}} a_{k_3} a_{k_{3'}}$\\ \hline
 $T_2$ & $-+-+$ &$\bar u(-k_1)(\gamma_\mu {q_2}^\mu + m) u(k_{2'})$ & $b_{-k_1}^\dagger b_{k_{2'}} a_{-k_3}^\dagger a_{k_{3'}}$\\ \hline
 $T_3$ & $-++-$ &$\bar u(-k_1)(\gamma_\mu {q_2}^\mu + m) u(k_{2'})$ & $b_{-k_1}^\dagger b_{k_{2'}} a_{k_3} a_{-k_{3'}}^\dagger$\\ \hline
 $T_4$ & $-+--$ &$\bar u(-k_1)(\gamma_\mu {q_2}^\mu + m) u(k_{2'})$ & $b_{-k_1}^\dagger b_{k_{2'}} a_{-k_3}^\dagger a_{k_{3'}}^\dagger$\\ \hline
 $T_5$ & $+-++$ &$\bar v(k_1)(\gamma_\mu {q_2}^\mu + m) v(-k_{2'})$ & $d_{k_1} d_{-k_{2'}}^\dagger a_{k_3} a_{k_{3'}}$\\ \hline
 $T_6$ & $+--+$ &$\bar v(k_1)(\gamma_\mu {q_2}^\mu + m) v(-k_{2'})$ & $d_{k_1} d_{-k_{2'}}^\dagger a_{-k_3}^\dagger a_{k_{3'}}$\\ \hline
 $T_7$ & $+-+-$ &$\bar v(k_1)(\gamma_\mu {q_2}^\mu + m) v(-k_{2'})$ & $d_{k_1} d_{-k_{2'}}^\dagger a_{k_3} a_{-k_{3'}}^\dagger$\\ \hline
 $T_8$ & $+---$ &$\bar v(k_1)(\gamma_\mu {q_2}^\mu + m) v(-k_{2'})$ & $d_{k_1} d_{-k_{2'}}^\dagger a_{-k_3}^\dagger a_{-k_{3'}}^\dagger$\\ \hline
 $T_9$ & $ -++-$ & $\bar u(-k_1) (\gamma_\mu {q_2}^\mu + m) v(-k_{2'})$ & $b_{-k_1}^\dagger d_{-k_{2'}}^\dagger a_{k_3} a_{k_{3'}}$ \\
 \hline
 $T_{10}$ & $ --+-$ & $\bar u(-k_1) (\gamma_\mu {q_2}^\mu + m) v(-k_{2'})$ & $b_{-k_1}^\dagger d_{-k_{2'}}^\dagger a_{-k_{3'}}^\dagger a_{k_3}$ \\
 \hline
 $T_{11}$ & $ -+--$ & $\bar u(-k_1) (\gamma_\mu {q_2}^\mu + m) v(-k_{2'})$ & $b_{-k_1}^\dagger d_{-k_{2'}}^\dagger a_{k_3} a_{-k_{3'}}^\dagger$ \\
 \hline
 $T_{12}$ & $+-++$ & $\bar v(k_{1}) (\gamma_\mu {q_2}^\mu + m) u(k_{2'})$ & $ d_{k_{1}} b_{k_{2'}} a_{-k_3}^\dagger a_{k_{3'}}$\\ \hline
 $T_{13}$ & $++-+$ & $\bar v(k_{1}) (\gamma_\mu {q_2}^\mu + m) u(k_{2'})$ & $ d_{k_{1}} b_{k_{2'}} a_{k_3} a_{-k_{3'}}^\dagger$\\ \hline
 $T_{14}$ & $+--+$ & $\bar v(k_{1}) (\gamma_\mu {q_2}^\mu + m) u(k_{2'})$ & $ d_{k_{1}} b_{k_{2'}} a_{-k_2}^\dagger a_{-k_{3'}}^\dagger$\\ \hline
 
\end{tabular}
  \caption{\justifying \textbf{Discrete} $H_\text{fe}(\lambda)$: $k_1, k_{2'} \in \mathbb{Z} + 1/2$, $k_3, k_{3'} \in \mathbb{Z} \backslash \{0\}$. The column $\theta$ is simplified as $\theta \equiv \theta(k_1, k_{2'}, k_3, k_{3'})$ for brevity. The spinors are implicitly assumed to mean $u(k) = u(q^+(k))$.
  }
  \label{table:Hfe}
\end{table}

The discretized fermion exchange term $H_{\text{fe}}(\lambda)$ is given as
\begin{align}
\label{eq:Hfe-discrete}
    H_{\text{fe}}(\lambda) &= \frac{1}{2L(4\pi)^4}\sum_n \sum\limits_{\substack{k_1, k_2, k_3\\k_{1'}, k_{2'}}}\frac{\delta_{k_1, -k_2-k_3}\delta_{k_2, k_{2'} + k_{3'}}}{\sqrt{|k_1k_{2'}k_{3}k_{3'}|}}\nonumber \\
    &\times \mathcal{R}\Bigg(\lambda; \Big(q^-(k_1) + q^-(k_2) + q^-(k_3)\Big), \nonumber\\
    &\hspace{2cm}\Big(q^-(k_{2'}) +  q^-(k_{3'}) - q^-(k_2)\Big)\Bigg)\nonumber\\&\times :T^{(\mathrm{IV})}_n(k_1,k_{2'}, k_3, k_{3'}):,
\end{align}
where $k_1, k_2,k_{2'}$ correspond to the fermion momenta, while $k_3$ and $k_{3'}$ correspond to the boson momenta.  $T^{(\mathrm{IV})}_n$ corresponds to a term in a row in Table~\ref{table:Hfe}.

The discretized loops and counterterms can be combined, giving a renormalized term, independent of the regulator. The forms are the same as the corresponding free discrete terms in Eq.~\ref{eq:H-free-discrete}, but modified by the effective loop integrals, $m^2(\lambda;k)$ and $\mu^2(\lambda;k)$. The loop integrals are finite since $\lambda$ is finite, and they don't depend on $\Lambda$. While it may be possible to find closed forms of these integrals, they can be computed numerically.
\begin{align}
\label{eq:effective-fermion-mass}
    H_{\delta m^2}(\lambda) &+ X_{\delta m^2}=\nonumber \\
    &\frac{L}{2\pi}\sum_{k \in \mathbb{Z}^+ - 1/2} \Bigg(\frac{m^2(\lambda;k)}{k}\Bigg) \Big(b_{k}^\dagger b_{k} + d_k^\dagger d_k \Big),
\end{align}
and
\begin{align}
\label{eq:effective-boson-mass}
    H_{\delta \mu^2}(\lambda) &+ X_{\delta \mu^2}=\nonumber \\
    &\frac{L}{2\pi}\sum_{k \in \mathbb{Z}^+} \Bigg(\frac{\mu^2(\lambda;k)}{k}\Bigg) a_k^\dagger a_k,
\end{align}
where 
\begin{align}
    m^2(\lambda;k) &=  \int_0^1 \frac{dx}{x(1-x)}\frac{1}{\mathcal{M}_{fb}^2(x) - m^2}\times\nonumber \\
    &\hspace{1cm}\frac{m^2(1 + x)^2}{x}e^{-2L^2 /(4\pi k^2\lambda^2)\Big( \mathcal{M}_{fb}^2(x) - m^2\Big)^2}
\end{align}
and 
\begin{align}
    \mu^2(\lambda;k) &= 2 \int_0^1 \frac{dx}{x(1-x)}\frac{1}{\mathcal{M}_{f\bar f}^2(x) - \mu^2}\times\nonumber \\
    &\frac{m^2(2x - 1)^2}{x (1-x)}e^{-2L^2 /(4\pi k^2\lambda^2)\Big( \mathcal{M}_{f \bar f}^2(x) - \mu^2\Big)^2}.
\end{align}
Again, $\mathcal{M}^2$ is the invariant mass of the two particles in the loop (see Eq.~\ref{eq:inv-mass}).



The discretized effective Hamiltonian is 
\begin{align}
\label{eq:discrete-effective-hamiltonian}
    H(\lambda) &= H_0 + H_{\bar \psi \psi \phi}(\lambda) + H_{\bar \psi \phi \phi \psi}\left(\lambda\right)\nonumber\\
    &+ H_{\textrm{fe}}(\lambda) +H_{\textrm{be}}(\lambda)\nonumber \\
    &+H_{\delta m^2}(\lambda) + X_{\delta m^2}+ H_{\delta \mu^2}(\lambda) + X_{\delta \mu^2}
\end{align}
coming from Eqns.~\ref{eq:H-free-discrete},~\ref{eq:effective-H-3-discrete},~\ref{eq:effective-H-4-discrete},~\ref{eq:Hbe-discrete},~\ref{eq:Hfe-discrete},~\ref{eq:effective-fermion-mass} and~\ref{eq:effective-boson-mass}.
This Hamiltonian serves as the input to all numerical calculations performed in this paper (apart from the bare Hamiltonian spectrum in Fig.~\ref{fig:bare_spectrum}).
\section{Numerical calculations of the $M^2$ spectrum}
\label{sec:spectrum}

Calculating the eigenvalues of Eq.~\ref{eq:1p1Deigenvalue} gives the spectrum of the corresponding bound eigenstates.
After choosing a harmonic resolution, $K$, a combinatorics problem arises, where one must construct all states whose discrete momentum quantum numbers $k$ sum to $K$. 
This collection of Fock states can now be partitioned into sectors by their constitution, i.e. $$ |\psi \rangle \in \Bigl\{|f\rangle, |\bar f\rangle, |b\rangle, |fb\rangle, |f\bar fb\rangle, \dots \Bigr\},$$
where $\ket{f}$ corresponds to all Fock states of a single fermion, $\ket{fb}$ corresponds to all Fock states of a fermion $+$ a boson, etc. such that their momenta sum to $K$.
The \textit{charge} $Q\equiv N_f - N_{\bar f}$ is a conserved quantity that is defined as the number of fermions minus the number of antifermions.

\begin{figure}
    \centering
    \makebox[\textwidth][l]{
        \hspace{-0.5cm}%
        \includegraphics[width=\linewidth]{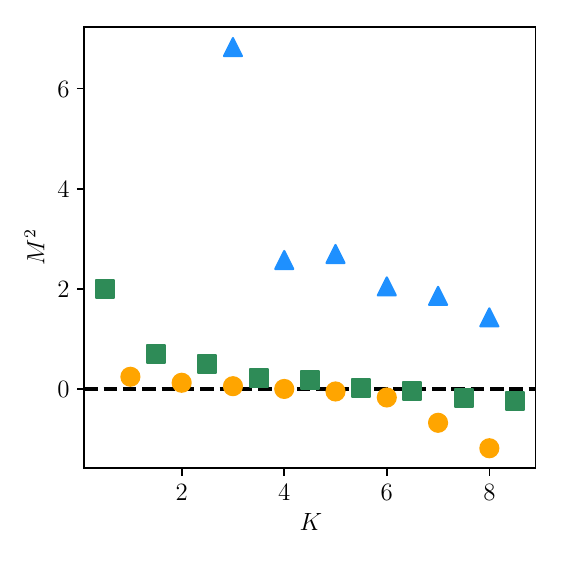}
    }\caption{\justifying\textbf{Bare spectrum} Canonical Yukawa Hamiltonian spectrum with increasing harmonic resolution. 
    Yellow circles correspond to the $Q = 0$ sector, green squares correspond to the $Q = 1$ sector, and blue triangles correspond to the $Q = 2$ sector. A cutoff on the maximum number of particles, $n_p$ in a Fock state is made ($n_p = 4$).
    The relevant parameters are $m = 1$, $\mu = 0.5$, $g = 1$.}
    \label{fig:bare_spectrum}
\end{figure}

Fig.~\ref{fig:bare_spectrum} shows that the lowest eigenvalues in the charge $Q = 0, 1,$ and $2$ sectors of the bare Hamiltonian tend towards negative $M^2$ values. 
This arises because the canonical Hamiltonian is ill-defined. 
This problem should be solved by a proper renormalization procedure. 

\begin{figure}
     \centering
     \begin{subfigure}[b]{0.45\textwidth}
         \centering
         \includegraphics[width=\textwidth]{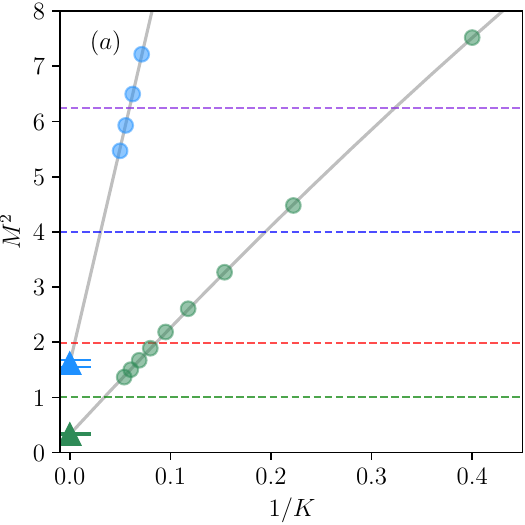}
         \label{fig:g_1}
     \end{subfigure}
     \hfill
     \begin{subfigure}[b]{0.45\textwidth}
         \centering
         \includegraphics[width=\textwidth]{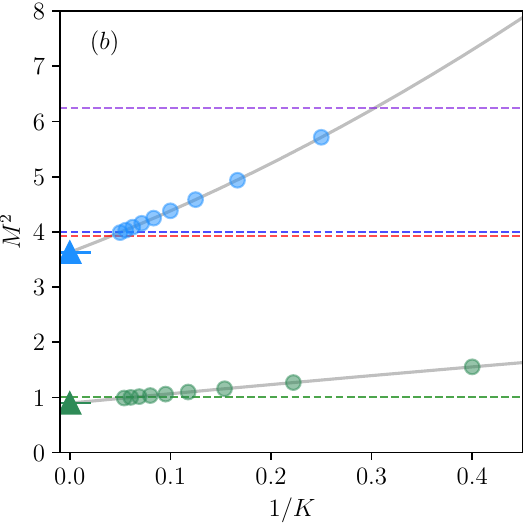}
         \label{fig:g_0.3}
     \end{subfigure}
        \caption{\justifying\justifying \textbf{Extrapolations} Convergent $M^2$ eigenvalues of the renormalized Hamiltonian. 
            Relevant parameters in both plots are $m = 1$, $\mu = 0.5$, and $\lambda = 10^{6}$. 
            The horizontal dashed lines represent, respectively from bottom to top, $m^2$, non-relativistic approximation, $\left(2m \right)^2$, and $\left(2m + \mu \right)^2$.
            The top set of dots corresponds to the lowest eigenvalue of the $ff$, $ffb$ sectors, while the bottom set of dots corresponds to the lowest eigenvalue of the $f$, $fb$ sectors.
            The fits are of the form $a + b/K + c/K^2$. (a) ($g = 1$)  $1/K \rightarrow 0$ extrapolations, yielding the values $M^2_f = 0.33m^2$, $M_{ff}^2 = 1.616m^2$ (compared to the non-relativistic value of $M_{ff}^2 = 1.998m^2$). (b) ($g = 0.3$) $1/K \rightarrow 0$ extrapolations, yielding the values $M^2_f = 0.896m^2$, $M_{ff}^2 = 3.628m^2$ (compared to the non-relativistic value of $M_{ff}^2 = 3.933m^2$). In both plots, there are error bars included on the triangles which correspond to the uncertainty in the fit.}
        \label{fig:renormalized}
\end{figure}

The discretized effective Hamiltonian in Eq. ~\ref{eq:discrete-effective-hamiltonian} can be numerically diagonalized in a Fock basis to give the renormalized spectrum, valid up to second order in the coupling.
Fig.~\ref{fig:renormalized} shows the extrapolated fermion-fermion (top dots) effective bound state $M^2$ extrapolated to infinite harmonic resolution for two different couplings: (a) $g = 1$ and (b) $g = 0.3$. 

The red dashed lines in the subplots of Fig.~\ref{fig:renormalized} are the lowest eigenvalue of the non-relativistic approximation of the 1D Yukawa potential:
\begin{equation}
\label{eq:V-yukawa}
    V(z) = \frac{-g^2}{2\mu}e^{-\mu|z|}.
\end{equation}
This potential admits a finite number of bound states with $E < 0$, depending on the choices of $g$ and $\mu$, and a continuum of scattering states with $E > 0$. For a derivation of Eq.~\ref{eq:V-yukawa}, see Appendix~\ref{sec:non-rel}. Finally, the eigenvalue of the non-relativistic 1D Yukawa potential in equal time, $E$, can be mapped~\cite{serafin2019bound} to the light-front eigenvalue, $M^2$, by:
\begin{equation}
    E = \frac{M^2 - \left(m_1 + m_2 \right)^2}{2\left(m_1 + m_2 \right)},
\end{equation}
where $m_1, m_2$ are the masses of the two particles.

The bottom (green) dots in both subplots in Fig.~\ref{fig:renormalized}, come from diagonalizing the Hamiltonian matrix in the Fock basis $$\Big\{\ket{f}, \ket{fb} \Big\}, $$ which are in the $Q = 1$ charge sector.
These points are included, not to show a bound state (as a single fermion bound state is not a well-defined object), but rather to show how well the renormalization condition has been satisfied.

The top (blue) dots in both subplots in Fig.~\ref{fig:renormalized} come from diagonalizing the Hamiltonian matrix in the Fock basis $$\Big\{\ket{ff}, \ket{ffb}\Big\},$$ which are in the $Q = 2$ sector.

\begin{figure}
    \centering
    \includegraphics[width=\linewidth]{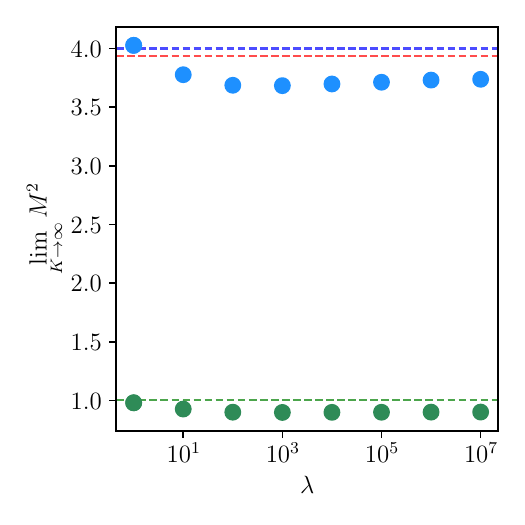}
    \caption{\justifying\textbf{Eigenstate scaling with $\lambda$} Extrapolated lowest $M^2$ values in $\{\ket{f}, \ket{fb}\}$ sector (bottom) and $\{\ket{ff}, \ket{ffb}\}$ sector (top) as a function of $\lambda$ at $g = 0.3$, $m = 1$, $\mu= 0.5$, and $K_{\textrm{max}} = 20$.
    As $\lambda$ is decreased to $\lambda \sim 1$, the interactions have been dampened so much that the approximation becomes the free theory.
    }
    \label{fig:flow}
\end{figure}

\begin{figure*}
    \centering
    \includegraphics[width = \linewidth]{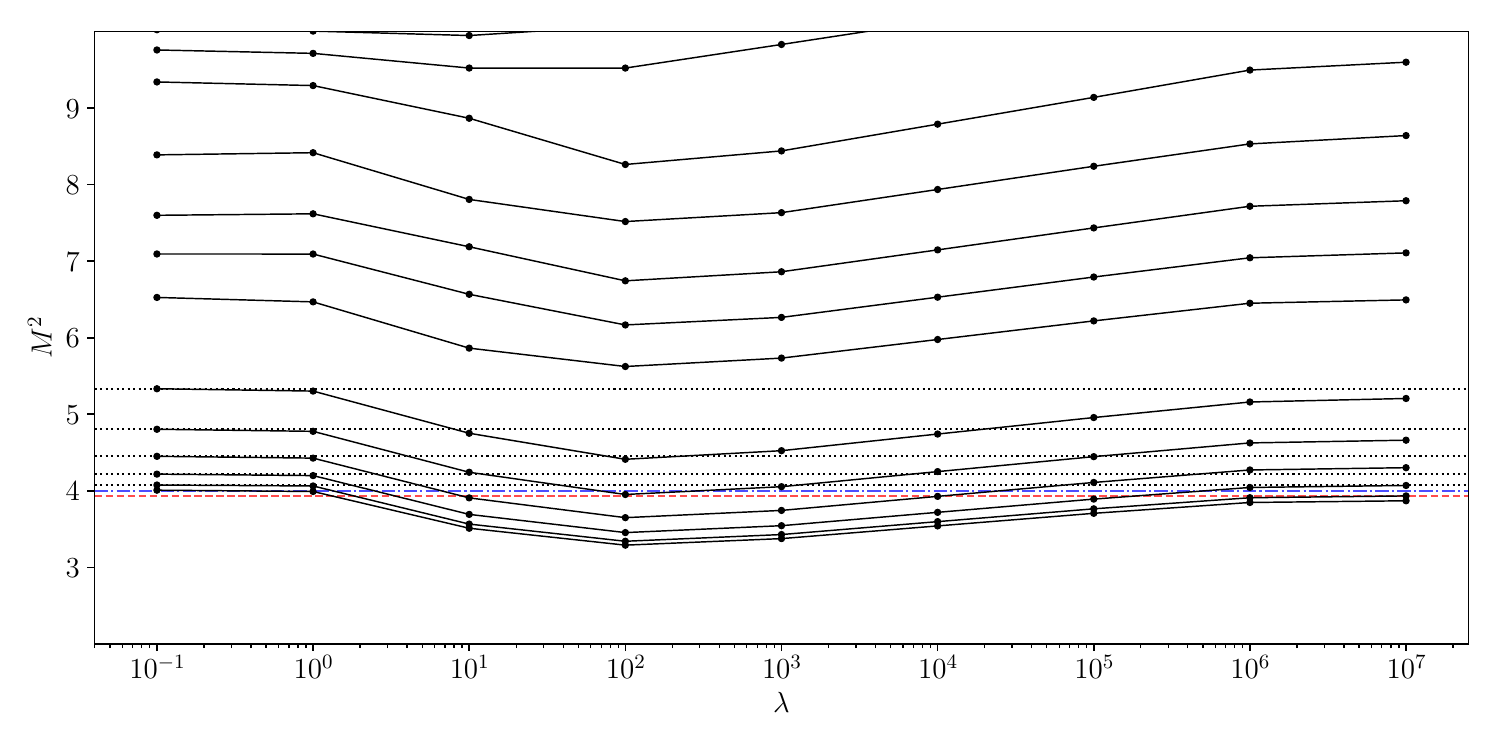}
    \caption{\justifying \textbf{Flow of Renormalized Hamiltonian Eigenvalues:} The Yukawa model parameters for this plot are $m = 1$, $\mu = 0.5$, $g = 0.5$, and $K = 22$. The Hamiltonian was diagonalized in the Fock basis: $\Big\{ \ket{ff}, \ket{ffb}, \ket{ffbb}, \ket{fff\bar f}\Big\}$. The blue (dashed and dotted) horizontal line corresponds to the mass of two free fermions, while the red (dashed) line is the non-relativistic approximation. The grey (dotted) horizontal lines correspond to the first few eigenvalues of the free ($g = 0$) Hamiltonian (above the $4m^2$ free Hamiltonian eigenvalue). At $\lambda \sim 1$, the calculated eigenvalues line up with the free theory, which is an incorrect approximation to the interacting theory. In this limit, the calculated eigenvalues match the free theory eigenvalues because of the form factors, which strongly dampen the interactions in this regime. Note that the higher spectrum is cutoff in order to show the behavior of the low-lying eigenstates more clearly.}
    \label{fig:flow-full-spectrum}
\end{figure*}

The `flow' of the Hamiltonian under the RGPEP equations is shown in Figures~\ref{fig:flow},~\ref{fig:flow-full-spectrum}. Fig.~\ref{fig:flow} shows the extrapolated ($K \rightarrow \infty$) lowest $M^2$ eigenvalues flowing with $\lambda$ for the same sectors in Fig.~\ref{fig:renormalized}. Fig.~\ref{fig:flow-full-spectrum} shows the flow of the entire spectrum with $\lambda$ for fixed $K = 22$, and in a larger Fock basis: $$\Big\{\ket{ff}, \ket{ffb}, \ket{ffbb}, \ket{fff\bar f} \Big\},$$ also in the $Q = 2$ sector.

The boost-invariance of the eigenvalues $M^2$ is only approximate due to our choice of generator in Eq.~\ref{eq:rgpep-generator}.
The form factors that arise from this generator, which modify the vertices, give explicit dependence on longitudinal momentum in the Hamiltonian matrix element.
Therefore, to obtain the same $M^2$ eigenvalue when $P^+$ is changed, one must rescale $\lambda \rightarrow P^+\lambda$.
In this sense, boost-invariance is restored. 
Other generators, such as the one used in Ref.~\cite{glazek2012perturbativeformulaerelativisticinteractions}, maintain boost-invariance; however, we choose the generator in Eq.~\ref{eq:rgpep-generator} because it leads to well-defined matrix elements in the color-singlet subspace when studying QCD~\cite{serafin2024dynamics, serafin-yukawa}.


\begin{figure}
    \centering
    \includegraphics[width = \linewidth]{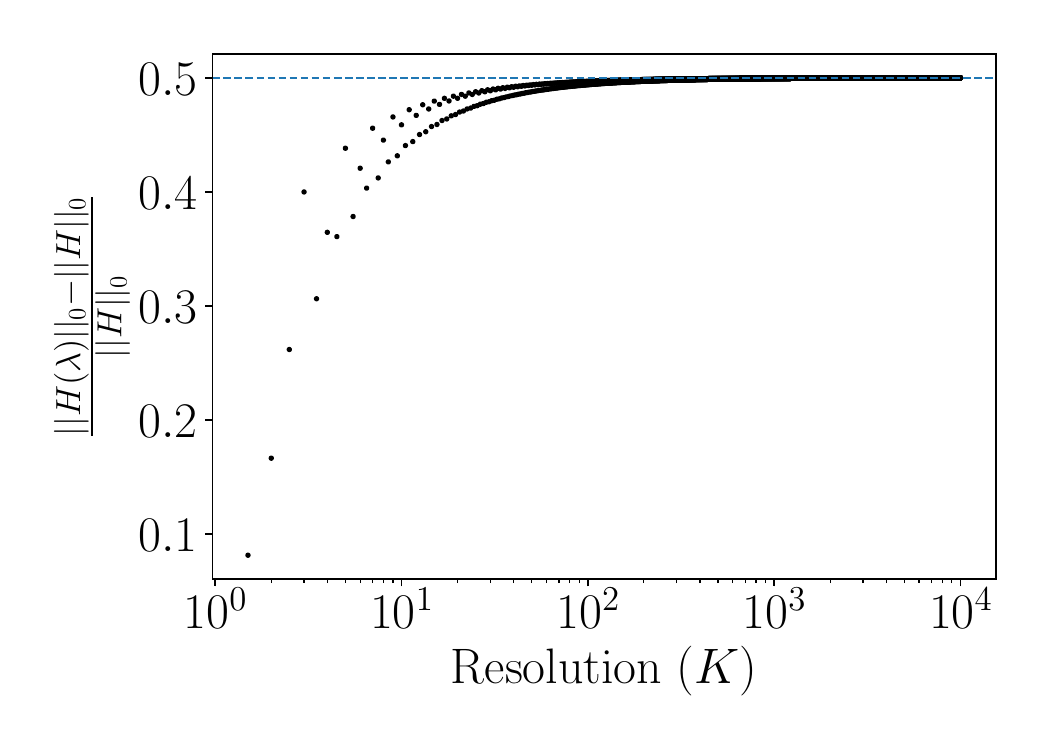}
    \caption{\justifying\textbf{Hamiltonian scalings} Relative scaling of the number of terms in the effective Hamiltonian $||H(\lambda)||_0$ vs. the number of terms in the bare Hamiltonian $||H||_0$. The scaling behaves as $\sim 0.5 + o(1)$, where $o(1) \rightarrow 0$ as $K \rightarrow \infty$.}
    \label{fig:n_terms}
\end{figure}

\begin{figure}
    \centering
    \includegraphics[width=\columnwidth]{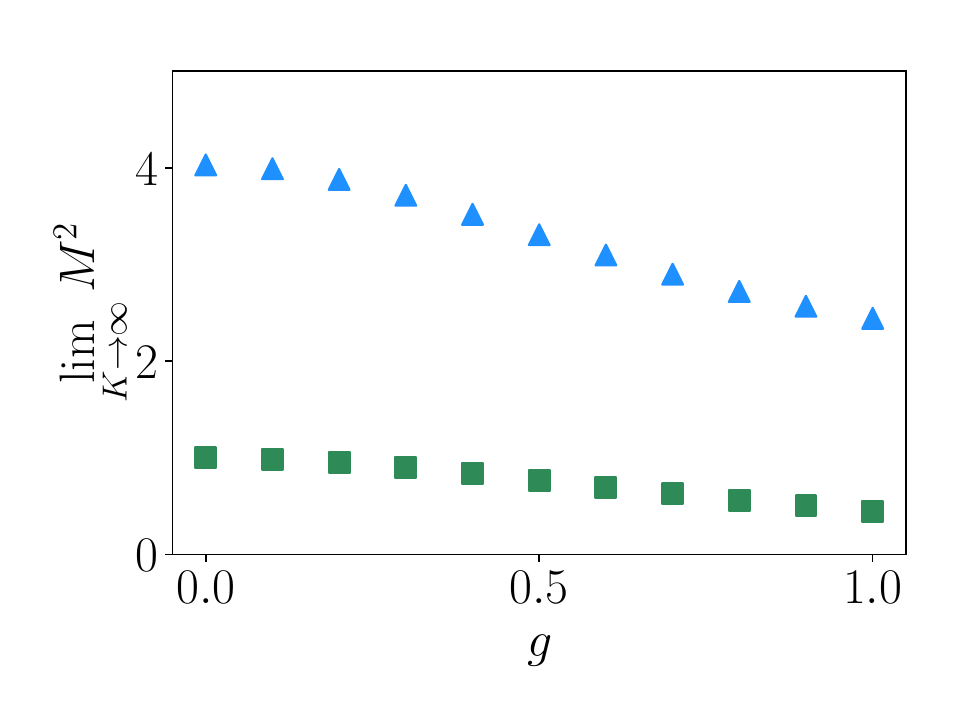}
    \caption{\justifying \textbf{Extrapolated $M^2$ vs. $g$:} As $g$ is increased, the extrapolated $M^2$ eigenvalue begins to decrease. This shows the perturbative solution to the RGPEP equation breaking down at higher $g$. In this plot, $K_{\textrm{max}} = 20$,  $m = 1, \mu = 0.5$. The squares and triangles describe the same sectors as Fig.~\ref{fig:bare_spectrum}. }
    \label{fig:M2-vs-g}
\end{figure}

\begin{figure*}
    \centering
    \includegraphics[width=\linewidth]{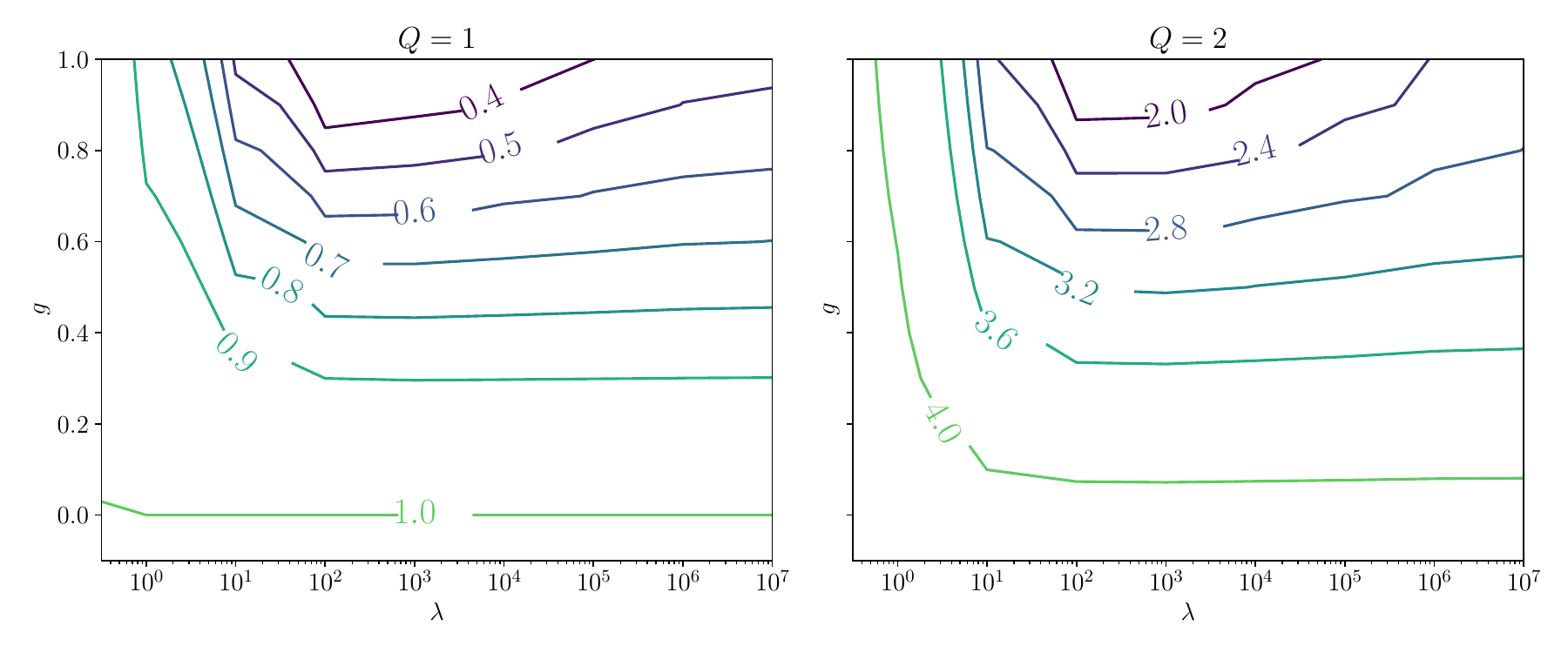}
    \caption{\justifying \textbf{Eigenvalue scaling with both $g$ and $\lambda$:} Contour plots displaying extrapolated $M^2(\lambda, g)$ eigenvalues at $K_{\textrm{max}} = 20$, $m = 1, \mu = 0.5$. The Fock basis used is the same as that in Fig.~\ref{fig:renormalized}. At small $g$, it is clear that the similarity transformation maintains the same spectrum, whereas as $g$ increases, this begins to break down.}
    \label{fig:m2-vs-g-lambda}
\end{figure*}

The scaling of the lowest $M^2$ eigenvalue in the $Q = 1$ and $Q = 2$ sectors (in the same basis in Fig.~\ref{fig:renormalized}) vs. the coupling constant $g$ is shown in Fig.~\ref{fig:M2-vs-g}.
The combined behavior of $M^2(g, \lambda)$ is shown in Fig.~\ref{fig:m2-vs-g-lambda}.

An important metric that describes the renormalized Hamiltonian is the number of terms.
The number of terms in the renormalized Hamiltonian, denoted by $||H(\lambda)||_0$, is greater than the number of terms in the bare Hamiltonian, denoted $||H||_0$.
Fig.~\ref{fig:n_terms} shows the relative scaling in the number of terms in the bare vs. effective Hamiltonians.

Asymptotically (at $K \rightarrow \infty$), the scaling for the bare Hamiltonian is:
\begin{equation}
    ||H||_0 = \frac53K^3 + \mathcal{O}(K^2),
\end{equation}
while the scaling for the renormalized Hamiltonian is:
\begin{equation}
    ||H(\lambda)||_0 = \frac{5}{2}K^3 + \mathcal{O}(K^2).
\end{equation}
Calculating the relative increase in the number of terms in the effective Hamiltonian compared to the bare Hamiltonian, we compute $$\frac{||H(\lambda)||_0 - ||H||_0}{||H||_0},$$ which tends to a constant factor of $1/2$ as $K \rightarrow \infty$, shown in Fig.~\ref{fig:n_terms}. This implies there are roughly $50\%$ more terms asymptotically in the effective Hamiltonian compared to the bare Hamiltonian.

For weak coupling, we expect the two fermion bound state mass to be less than the invariant mass of two free fermions $4m^2$.
In both cases of coupling tested in this paper, this is true.
Additionally, the ground state masses we obtain (for small coupling) qualitatively agree with comparable results from 3+1D calculations~\cite{Qian:2020ttn, PhysRevD.47.1599}, in that they elicit stronger binding than those produced by the non-relativistic results.

\section{Parton Distribution Functions}
\label{sec:pdfs}

A \textit{parton} is a constituent of a bound state of any quantum field theory.
Parton distribution functions~\cite{Collins_2011, Soper_1997} for fermionic fields are given by:
\begin{equation}
    \label{eq:qcd-pdf}
    f_i(x) = \int \frac{dy^-}{4\pi}e^{ixP^+y^-}\bra{P}\bar \psi_i(y^-)\gamma^+\psi_i(0)\ket{P}.
\end{equation}
These functions contain information about the distribution of longitudinal momentum carried by the constituents that make up a particular bound state, described by $\ket{P}$. 

Given a bound state $\ket{\psi}$, its parton distribution function can be determined by calculating

\begin{equation}
    f_p(x) \equiv \matrixel*{\psi}{\hat N_p(x)}{\psi},
\end{equation}
where $\hat N_p(x)$ is given as the number operator of a particle of type $p$ (i.e. $p = $ fermion, antifermion, boson) with momentum fraction $x = k/K$~\cite{Hornbostel}.
Increasing $K$ increases the number of points along the $x$-axis in a plot of a parton distribution function, hence why it is referred to as the harmonic resolution.

\begin{figure}
    \centering
    \makebox[\textwidth][l]{
        \hspace{-0.41cm}%
        \includegraphics[width=\linewidth]{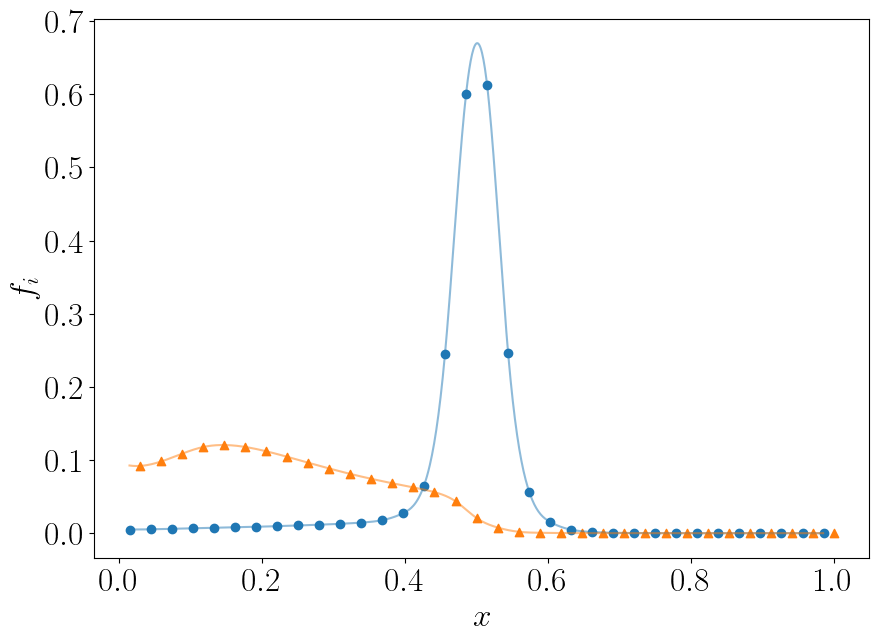}
    }
    \caption{\justifying\textbf{Parton Distribution Functions} (PDF) for the lowest $M^2$ bound state in the $\{\ket{ff}, \ket{ffb} \}$ sectors, with $f_i = f_f, 10f_b$. 
    The blue dots represent the fermionic PDF.
    The orange triangles represent the bosonic PDF, which is scaled up by a factor of 10 to show its behavior in the low-$x$ region.
    The solid lines are numerical fits to the data.
    The relevant parameters are $m = 1$, $\mu = 0.5$, $g = 1$, $\lambda = 10^{6}$, $K = 34$.
    }
    \label{fig:pdf}
\end{figure}

For example, the fermionic parton distribution function requires computing expectation values of a bound state $\ket{\psi}$ with the operator
\begin{equation}
    \hat N_f(x) = b_k^\dagger b_k.
\end{equation}
Fig.~\ref{fig:pdf} gives the parton distribution function calculated for the state with the lowest $M^2$ eigenvalue in the $$\Bigl\{\ket{ff}, \ket{ffb} \Bigr\}$$ sectors.

The main features of this plot are that the fermionic parton distribution function peaks at around $x = 0.5$, implying that each fermion shares roughly half of the total momentum. Additionally, there is a small bosonic contribution from the $\ket{ffb}$ states that arises at small $x$ values.

\section{Block-Encoding Metrics}
\label{be-metrics}

Renormalization introduces additional terms into the Hamiltonian, and may change its norm, which can affect the cost of quantum simulation algorithms.
In this Section, we evaluate the relative cost of simulating the renormalized Hamiltonian, as compared to the bare Hamiltonian, to evaluate the effect of renormalization on the cost of quantum simulation. 

Many Hamiltonian simulation algorithms are built upon block-encodings of the Hamiltonian, therefore we quantify the cost of quantum simulation by calculating the quantum resource estimates for implementing a single block-encoding of the Yukawa Hamiltonian.
The block-encodings for each Hamiltonian are constructed using the Ladder Operator Block-Encoding (LOBE) framework~\cite{lobe}. 

Block-encoding~\cite{childs, chakraborty, lin} is a subroutine that provides access to information regarding non-unitary operators, such as the Hamiltonian, within quantum algorithms.
Block-encodings can be used for various methods including estimating the eigenvalues of the block-encoded operator, generating the time-evolution operator, or producing arbitrary polynomial functions of the time-evolution operator~\cite{poulin, gilyen_qsvt, grand-unification, linear-T, qubitization, qsp, childs-lcu}.

Quantum resource estimates for a block-encoding can be quantified in terms of the number of qubits (space) and the number of operations (time) needed to implement the block-encoding.
Block-encodings require ancillae qubits, some of which are only temporarily allocated.
Therefore, we quantify the number of qubits needed in terms of both the number of block-encoding ancillae and the maximum number of qubits used.
The most costly operations to perform in fault-tolerant architectures are typically non-transversal operations.
Under the surface code~\cite{surface-code}, the $T$ gate and non-Clifford rotations about the $z$-axis are non-transversal, therefore we count these two operations to quantify the time complexity.
Finally, the overall rescaling factor associated with each block-encoding often proportionally increases the cost of implementing simulation algorithms constructed using block-encodings.
To account for this, we directly compare the rescaling factors associated with each Hamiltonian.


\begin{figure*}
    \includegraphics[width = \linewidth]{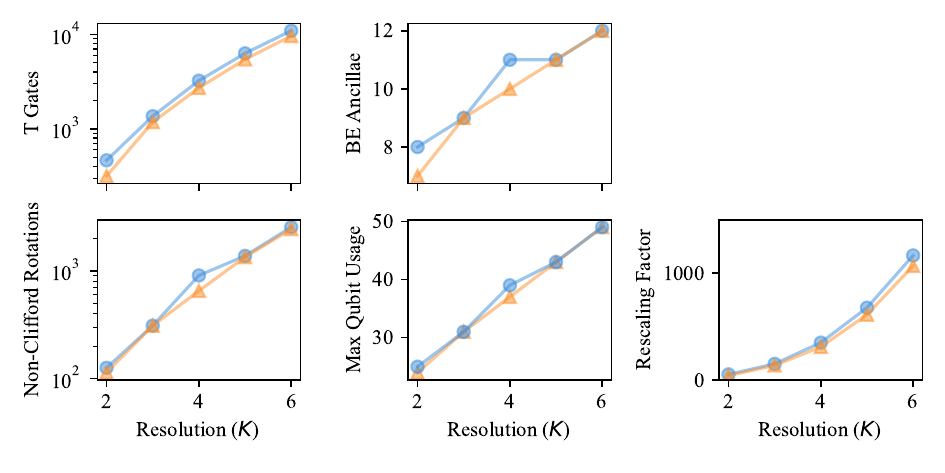}
    \caption{\justifying
        \textbf{Block-Encoding Metrics}
        The number of $T$ gates (upper-left), number of non-Clifford rotations (lower-left), block-encoding ancillae (upper-middle), maximum number of qubits used (lower-middle), and rescaling factor (lower-right) are shown as a function of the harmonic resolution ($K$).
        The renormalization parameter is set to $\lambda = 10^{7/2}$, the bosonic cutoff is fixed to $\Omega = 3$, and the coefficients $m$, $\mu$, and $g$ are set to $1$ for all data points.
        Metrics for the bare Hamiltonians are shown in orange and the metrics associated with the renormalized Hamiltonian are shown in blue.
    }
    \label{fig:be-metrics}
\end{figure*}
\begin{figure*}
    \includegraphics[width = \linewidth]{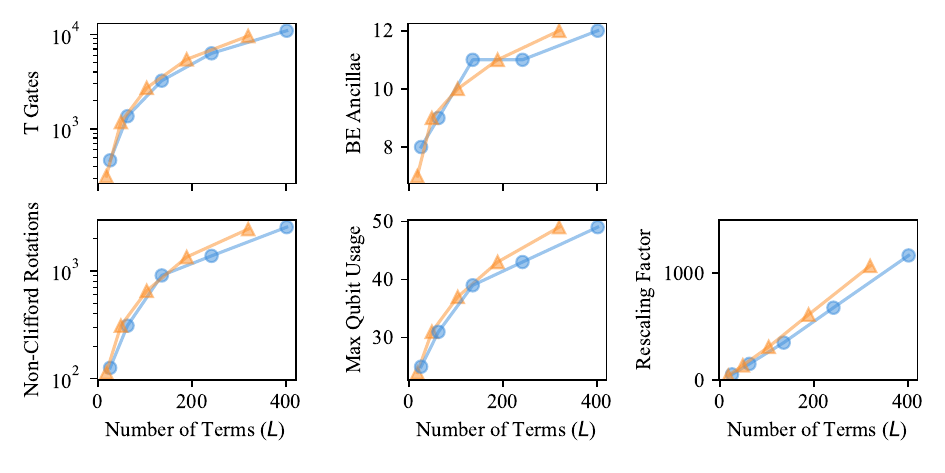}
    \caption{\justifying
        \textbf{Block-Encoding Metrics}
        The number of $T$ gates (upper-left), number of non-Clifford rotations (lower-left), block-encoding ancillae (upper-middle), maximum number of qubits used (lower-middle), and rescaling factor (lower-right) are shown as a function of the number of terms in a given Hamiltonian.
        The renormalization parameter is set to $\lambda = 10^{7/2}$, the bosonic cutoff is fixed to $\Omega = 3$, and the coefficients $m$, $\mu$, and $g$ are set to $1$ for all data points.
        Metrics for the bare Hamiltonians are shown in orange and the metrics associated with the renormalized Hamiltonian are shown in blue.
    }
    \label{fig:be-metrics-n-terms}
\end{figure*}

The quantum resource estimates are shown as a function of the harmonic resolution ($K$) in Fig.~\ref{fig:be-metrics} and as a function of the number of terms in the Hamiltonian in Fig.~\ref{fig:be-metrics-n-terms}.
The costs for the bare Hamiltonians are shown in orange and the costs for the renormalized Hamiltonians are shown in blue.

At each resolution, the quantum resource estimates for block-encoding the renormalized Hamiltonians are larger for all metrics that we consider.
However, the increase in quantum resources is a small portion of the overall cost.
As shown in Fig.~\ref{fig:be-metrics-n-terms}, the renormalized Hamiltonians require fewer quantum resources per term.
This implies that the increase in quantum resources is due to renormalization introducing more terms, but it does not introduce terms that are more expensive to block-encode on average.
These numerical quantum resource estimates, along with the results of Fig.~\ref{fig:n_terms}, suggest that renormalization up to second order increases the cost of quantum simulation algorithms that use block-encodings of the Yukawa Hamiltonian, in the worst case, by roughly $50\%$ asymptotically.

\section{Conclusions}
\label{sec:conclusions}

We give the effective Hamiltonian up to $\mathcal{O}\left(g^2\right)$ using the Renormalization Group Procedure for Effective Particles~\cite{serafin-yukawa}.
Both the bare and renormalized Hamiltonians are discretized via discretized light cone quantization. With the discrete Hamiltonians, we numerically compute mass spectra, and show that the counterterms added in the effective Hamiltonian lead to finite observables. The calculations done in this work are the most sophisticated RGPEP calculations done to date. 

Additionally, we have calculated the first estimates of quantum resources necessary for quantum simulation of a renormalized Quantum Field Theory, the Yukawa model in particular, in the front form.
As a metric for the number of quantum resources necessary for quantum simulation, we calculated the cost associated with a single block encoding of the Hamiltonian using the Ladder Operator Block Encoding framework~\cite{lobe}.

The results presented in Section~\ref{be-metrics} provide motivation that leads us to believe that quantum simulation of renormalized quantum field theories in the front-form approach is not an intractable framework. Other Hamiltonians must be studied to confirm this.

Even though state-of-the-art calculations in nuclear physics use chiral effective theory~\cite{WEINBERG1990288, KAPLAN1998390}, Yukawa theory as a model of the strong nuclear force has historical significance, offers simplicity, and is a natural bridge to QCD. 
It is for this reason the Yukawa model was studied in this work.

Future work can take many different forms.
One can study larger problems in different ways: increasing the number of Fock sectors, doing calculations at higher harmonic resolution, and considering the transverse $\vec{x}^\perp$ contributions using basis light-front quantization.
Additionally, calculating solutions to the RGPEP equation at $\mathcal{O}\left(g^3\right)$, should give both a more accurate spectrum than this paper offers, as well as allowing for the coupling constant, $g$, to be renormalized.

Another future direction is studying these techniques with a gauge theory (unlike Yukawa theory), which will necessarily present new issues. The simplest gauge theory that could be studied is the Schwinger model~\cite{schwinger}, or one could choose $SU(2)$ Yang-Mills theory~\cite{Peskin:1995ev}, which exhibits asymptotic freedom. If the main goal of simulating quantum field theories is to obtain results that can be experimentally verified with hadronic measurements, full 3 + 1D Quantum Chromodynamics~\cite{Peskin:1995ev} must be studied. We leave these many interesting directions for future work.

\section{Acknowledgements}
\label{sec:acknowledgement}
We thank James Vary and Yang Li for discussions on their previous work on the Yukawa model.
We thank Jose Emmanuel Flores for sharing code to calculate eigenvalues of the non-relativistic approximation. Carter M. Gustin, Kamil Serafin, Gary R. Goldstein and Peter J. Love are supported by US DOE Grant DE-SC0023707 under the Office of Nuclear Physics Quantum Horizons program for the “\textbf{Nu}clei and \textbf{Ha}drons with
\textbf{Q}uantum computers (NuHaQ)” project.
William A. Simon is supported by the Department of Defense (DoD) through the National Defense Science \& Engineering Graduate (NDSEG) Fellowship Program.
Alexis Ralli was supported by the STAQ project under award NSF-PHY-232580.

\bibliography{main}

\appendix

\section{Non-relativistic Yukawa Theory in 1D}
\label{sec:non-rel}

All non-relativistic values quoted in this work come from the following transformation from the instant time eigenvalue $E$ to the light-front eigenvalue $M^2$~\cite{serafin2019bound}:

\begin{equation}
    E = \frac{M^2 - \left(m_1 + m_2 \right)^2}{2\left(m_1 + m_2 \right)},
\end{equation}
where $m_1, m_2$ are the masses of the scattering or bound particles.

Understanding the non-relativistic limits of a theory of interest is important as it gives a guide as to roughly what solutions one should expect. 
The non-relativistic limit can be studied by taking $\left(p - p' \right)^2 = -|\textbf{p} - \textbf{p'}|^2 + \mathcal{O}\left(\textbf{p}\right)^4$ in the scattering matrix element~\cite{Peskin:1995ev}.
For the \textit{t}-channel interaction, the (spin-less) scattering matrix element is 

\begin{align}
    \label{eq:iM}
    i\mathcal{M} &= \bar u(p') u(p) \frac{-ig^2}{\left(p - p' \right)^2 - \mu^2}\bar u(k') u(k) \nonumber \\
                 &= \frac{(2m)^2ig^2}{|\textbf{p} - \textbf{p'}|^2 + \mu^2}.
\end{align}

The Born approximation to the scattering amplitude in non-relativistic quantum mechanics is given as $\matrixelement{p'}{iT}{p} = -i\tilde V(\textbf{q})(2\pi)\delta(E_p - E_{p'})$. 
Defining $\textbf{q} \equiv \textbf{p} - \textbf{p'}$, one can compare this expression with Eq.~\ref{eq:iM} to get 
\begin{equation}
    \tilde V(\textbf{q}) = \frac{-g^2}{|\textbf{q}|^2 + \mu^2},
\end{equation}
the 1D Yukawa potential in momentum space.
Note that the relativistic normalization that gives the two factors of $2m$ must be removed in the non-relativistic limit, as non-relativistic normalization must be imposed. 
Additionally, the delta disappears when integrating over the momentum of the target \cite{Peskin:1995ev}.

Taking a Fourier transform to position space (and un-bolding $q$ since it is a 1D vector), 

\begin{align}
    V(z) &= \int_{-\infty}^{\infty}\frac{dq}{2\pi}\tilde V(q) e^{iqz} \nonumber \\
         &= \frac{-g^2}{2\pi}\int_{-\infty}^{\infty}dq \frac{e^{iqz}}{q^2 + \mu^2}\nonumber \\
         &= \frac{-g^2}{2\pi}\int_{-\infty}^{\infty}dq \frac{e^{iqz}}{\left(q +i\mu \right)\left(q -i\mu \right)}
\end{align}

This integral must be handled separately for $z \geq 0$ and $z < 0$.
For $z \geq 0$, a counterclockwise contour is taken to enclose the pole at $q = i\mu$:

\begin{align}
    V(z) &= \frac{-g^2}{2\pi}2\pi i \frac{e^{i(i\mu)z}}{i\mu + i\mu}\nonumber \\
    &= \frac{-g^2}{2\mu}e^{-\mu z}\nonumber
\end{align}

For $z < 0$, a clockwise contour is taken to enclose the pole at $q = -i\mu$:
\begin{align}
    V(z) &= (-1)\frac{-g^2}{2\pi}2\pi i \frac{e^{i(-i\mu)z}}{-i\mu - i\mu}\nonumber \\
    &= \frac{-g^2}{2\mu}e^{\mu z}\nonumber
\end{align}

Both solutions can be taken together to give 

\begin{equation}
    V(z) = \frac{-g^2}{2\mu}e^{-\mu|z|}
\end{equation}

\begin{figure}[h]
    \centering
    \includegraphics[width = 9cm]{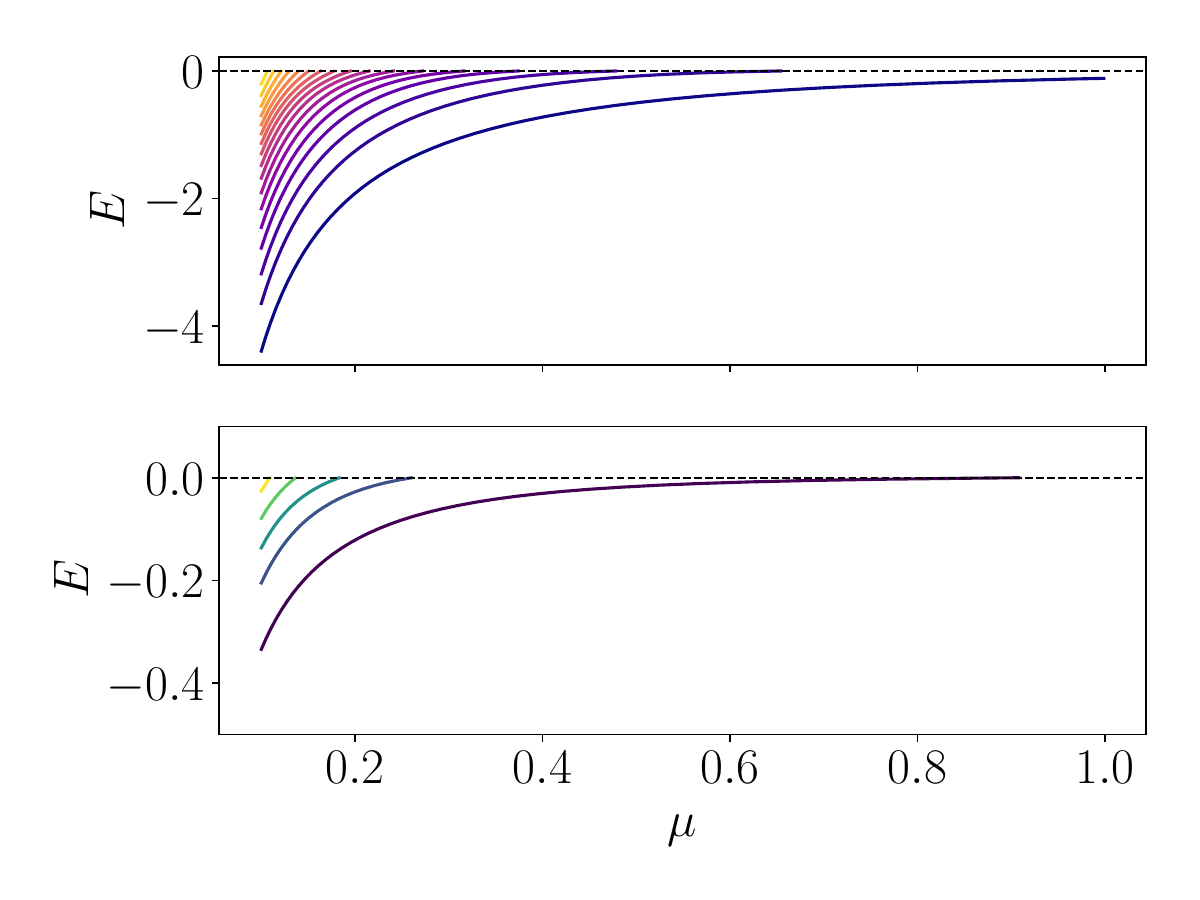}
    \caption{\justifying
        \textbf{Eigenvalues of the non-relativistic 1D Yukawa Potential} with changing boson mass $\mu$.
        (Top) $g = 1$.
        (Bottom) $g = 0.3$.
    }
    \label{fig:non-rel}
\end{figure}

Fig.~\ref{fig:non-rel} shows the eigenvalues at fixed $g$ while varying $\mu$. 
For larger $g$, one expects the number of bound states to increase. 
The limit $\mu \rightarrow 0$ doesn't recover the 1D Coulomb potential, as one gets in 3D.

Fig.~\ref{fig:non-rel} shows that for the plots in figure~\ref{fig:renormalized}, one would expect 1 (2) bound state(s) for $g = 0.3$ ($g = 1$) given the choice of $\mu = 0.5$.
Fig.~\ref{fig:pdf} shows a bound state.

\section{Solution to RGPEP Equation up to $\mathcal{O}\left(g^2\right)$ }
\label{sec:orders}

The RGPEP equation, Eq.~\ref{eq:rgpep}, is in general too difficult to solve exactly. A perturbative solution is needed in which the Hamiltonian is written as an expansion in powers of the coupling constant:
\begin{equation}
    \label{eq:H-expansion}
    H(s) = H_0(s) + gH^{(1)}(s) + g^2 H^{(2)}(s) + \mathcal{O}(g^3).
\end{equation}
This leads to equations at various orders in the coupling constant in Eqns.~\ref{eq:rgpep-order-by-order-a},~\ref{eq:rgpep-order-by-order-b}, and~\ref{eq:rgpep-order-by-order-c}.

\subsection{$\mathcal{O}\left(g^0\right)$ Solution}

At $\mathcal{O}\left(g^0\right)$, the RGPEP equation is 
\begin{equation}
\label{eq:H0-rgpep}
    \frac{dH^{(0)}(s)}{ds} = 0 \implies H^{(0)}(s) = H_0.
\end{equation}
Since there is no $s$-dependence in $H_0(s)$, this solution is the same in terms of $\lambda$. 
That is, 
\begin{equation}
    H^{(0)}(\lambda) = H_0
\end{equation}

This says that the free terms in the Hamiltonian are the same in the bare and effective Hamiltonians, i.e. they are not affected by the similarity transformation.
\subsection{$\mathcal{O}\left(g\right)$ Solution}
\label{sec:first-order}

At $\mathcal{O}\left(g\right)$, the RGPEP equation is 
\begin{equation}
\label{eq:first_order-appendix}
    \frac{dH^{(1)}(s)}{ds} = \Bigg[\Big[H_0, H^{(1)}(s)\Big], H_0\Bigg],
\end{equation}
where the zeroth order solution,~\ref{eq:H0-rgpep}, is explicitly used.

To solve Eq.~\ref{eq:first_order-appendix}, first define the generator at $\mathcal{O}\left(g\right)$ as 
\begin{equation}
    \label{eq:first-order-generator}
    \mathcal{G}^{(1)}(s) = \Big[H_0, H^{(1)}(s) \Big].
\end{equation}
It is necessary to define what $H^{(1)}(s)$ is. $H^{(1)}(s)$ is the $\mathcal{O}\left(g\right)$ terms in the Hamiltonian, i.e. Eq.~\ref{eq:H-free}. However, Eq.~\ref{eq:H-free} has no notion of the scaling parameter $s$. 

To solve Eq.~\ref{eq:first_order-appendix}, $H^{(1)}(s)$ is written as an ansatz of the same form as the standard 3-point interaction in Eq.~\ref{eq:H-3}, with a modification. We explicitly parameterize the interaction in Eq.~\ref{eq:H-3} by some arbitrary function $f(s;\{q\})$, where $\{q\}$ is a set of momentum variables in the interaction (e.g. $\{q\} = \{q_1, q_2, q_3\}$ for first order interactions), such that $H^{(1)}(s)$ takes the following form, with $f(s; \{q\})$ to be determined by solving Eq.~\ref{eq:first_order-appendix}. 
\begin{align}
    \label{eq:H-3-parameterized}
    H^{(1)}(s) = \int\limits_{1, 2, 3}&4\pi \delta(\mathcal{Q}^+)f(s;\{q\})r(1/\Lambda^2;\{q\})\nonumber \\
    &\times :\bar \psi(-q_1) \psi(q_2) \phi(q_3):.
\end{align}
Here, we have introduced a \textit{regulator}, $\Lambda$, via the use of a \textit{regulating function}, $r(1/\Lambda^2, \{q\})$, seen in Eq.~\ref{eq:H-3-parameterized}.

A regulator is a cutoff on some quantity that causes a singularity~\cite{Peskin:1995ev}. In light-front Hamiltonian field theory, divergences come from far off-diagonal matrix elements, corresponding to large energy differences at a vertex~\cite{serafin2019bound}. This motivates a regulator $\Lambda$ to cut-off, or dampen, matrix elements far from the diagonal. Many regulating functions $r(\Lambda, \{q\})$ can be chosen that accomplish this; however, the one chosen in this work is of the same form as the form factor:
\begin{equation}
\label{eq:regulator}
    r(\Lambda, \{q\}) = e^{-\left(q_1^- + q_2^- + q_3^- \right)^2/\Lambda^2}
\end{equation}
The reason for this should be evident soon.

It is now necessary to calculate 
\begin{align}
\label{eq:generator-first-order-comm}
    &\mathcal{G}^{(1)}(s) = \Bigg[  \frac{m^2}{2}\int \frac{dq^+}{4\pi} :\bar \psi(q) \frac{\gamma^+}{q^+}\psi(q):\nonumber \\
    &+ \frac{\mu^2}{2}\int \frac{dq^+}{4\pi} :\phi(-q)\phi(q):,
    \int\limits_{1,2,3} 4\pi \delta(\mathcal{Q}^+)f(s;\{q\})\nonumber \\
    &\hspace{2cm}\times r(\Lambda;\{q\}):\bar \psi(-q_1) \psi(q_2) \phi(q_3):\Bigg].
\end{align}
Working out the commutator in Eq.~\ref{eq:generator-first-order-comm}, one obtains 
\begin{align}
    \mathcal{G}^{(1)}(s) &= -\int\limits_{\substack{1,2,3}} 4\pi \delta(\mathcal{Q}^+) f(s;\{q\})r(\Lambda;\{q\})\nonumber \\
    &\times \mathcal{Q}^-:\bar \psi(-q_1) \psi(q_2) \phi(q_3):,
\end{align}
where $\mathcal{Q}^\pm = q_1^\pm + q_2^\pm + q_3^\pm$. This shows that the generator at $\mathcal{O}(g)$ is the same as $H^{(1)}(s)$, but with a factor of $\mathcal{Q}^-$ (sum of $q^-$ into the $H^{(1)}(s)$ interaction vertices) multiplied in front. To clean up notation, we write the generator as 
\begin{equation}
\label{eq:compact-generator}
    \mathcal{G}^{(1)}(s) = -\mathcal{Q}^- H^{(1)}(s),
\end{equation}
where it is assumed that the factor of $\mathcal{Q}^-$ is taken to be under the integral in the definition of $H^{(1)}(s)$.

Now one can work out the outer commutator on the right hand side of Eq.~\ref{eq:first_order-appendix}, with the new definition of the generator from Eq.~\ref{eq:compact-generator}.
\begin{align}
    \Bigg[\Big[H_0, H^{(1)}(s)\Big], H_0\Bigg] &= \Big[ \mathcal{G}^{(1)}(s), H_0\Big] \nonumber \\
    & =\Big[-\mathcal{Q}^-H^{(1)}(s), H_0 \Big] \nonumber \\
    & =-\mathcal{Q}^-\Big[H^{(1)}(s), H_0 \Big] \nonumber \\
    & = \mathcal{Q}^-\Big[H_0, H^{(1)}(s) \Big] \nonumber \\
    & = - \left(\mathcal{Q}^- \right)^2H^{(1)}(s).
\end{align}
This leads to the differential equation 
\begin{equation}
    \label{eq:first-order-compact}
    \frac{dH^{(1)}(s)}{ds} = - \left(\mathcal{Q}^- \right)^2H^{(1)}(s).
\end{equation}
Since only $f(s, \{q\})$ depends on $s$ in $H^{(1)}(s)$, the differential Eq.~\ref{eq:first-order-compact} reduces to
\begin{equation}
    \label{eq:first-order-f}
    \frac{df(s, \{q\})}{ds} = - \left(\mathcal{Q}^- \right)^2f(s, \{q\}).
\end{equation}
This leads to the form of the parameterization as   
\begin{equation}
    f(s;\{q\}) = e^{-s\left(\mathcal{Q}^-\right)^2} = e^{-s\left(q_1^- + q_2^- +     q_e^-\right)^2}.
\end{equation}
This is referred to as the \textit{form factor}, and is denoted $f(s; \mathcal{Q}^-)$. 

Note that in the main text of this manuscript, the form factor is written in terms of $\lambda$, not $s$.
It is implicitly assumed that the form factor in terms of $s$ is $\sim \exp(-s(\mathcal{Q}^-)^2)$, while in terms of $\lambda$ is $\sim \exp(-(\mathcal{Q^-}/\lambda)^2)$.
If one prefers, it may be useful to call the form factor in terms of $s$, $\tilde f(s; \mathcal{Q})$.
This distinction is purely conventional, and is the same as differentiating between $f(x) = 1/x$ or $f(y) = y$, where $y = 1/x$, which describe the same expression.

The full solution to Eq.~\ref{eq:first_order-appendix}, in terms of the physical $\lambda$ parameter is:
\begin{align}
    H^{(1)}(\lambda) = \int\limits_{1,2,3}4\pi &\delta(\mathcal{Q}^+)e^{-(q_1^- + q_2^- + q_3^-)^2/\lambda^2}r(\Lambda;\mathcal{Q}^-)\nonumber \\
    &\times:\bar \psi(-q_1) \psi(q_2) \phi(q_3):,
\end{align}
which matches Eq.~\ref{eq:first_order}.

\subsection{$\mathcal{O}\big(g^2\big)$ Solution}
At $\mathcal{O}\big(g^2\big)$, the RGPEP equation is
\begin{align}
\label{eq:second-order-appendix}
    \frac{dH^{(2)}(s)}{ds} = \Bigg[&\Big[H_0, H^{(2)}(s)\Big], H_0\Bigg] \nonumber \\
        &+ \Bigg[\Big[H_0, H^{(1)}(s)\Big],H^{(1)}(s)\Bigg].
\end{align}
We again assert an ansatz, $H^{(2)}(s)$ such that :
\begin{equation}
\label{eq:second-order-parameterized}
    H^{(2)}(s) = H_I(s) + H_{c}(s).
\end{equation}
$H_I(s)$ is a parameterized form of the second order term already present in the bare Hamiltonian such that $H_I(s) = F_1(s; \{q\})H_I$ (where again, the function $F_1$ is assumed to be under the integral in $H_I$). The regulator $\Lambda$ is assumed to be present in $H_I$. Again, we need to find an explicit form of $F_1(s;\{q\})$. 

There is an additional term present in Eq.~\ref{eq:second-order-parameterized} that was absent at the first order, that is the addition of $H_c(s)$, the \textit{contracted} terms. These terms arise from the second nested commutator in Eq.~\ref{eq:second-order-appendix}. The first nested commutator obtains contributions from both $H_I(s)$ and $H_{c}(s)$, while the second nested commutator will motivate the form of $H_c(s)$. Because of this, we will start by discussing this commutator. 

The second term on the right hand side of Eq.~\ref{eq:second-order-appendix} can be written as 
\begin{align}
\label{eq:commutator-contractions}
    \Big[\mathcal{G}^{(1)}(s), &H^{(1)}(s) \Big] = \Big(-\mathcal{Q}^-H^{(1)}(s)\Big)H^{(1)}(s)\nonumber \\
    &\hspace{1cm}- H^{(1)}(s)\Big(-\mathcal{Q}^-H^{(1)}(s)\Big) \nonumber \\
    &=:C\Big[\Big(-\mathcal{Q}^-H^{(1)}(s)\Big)H^{(1)}(s) \Big]:\nonumber \\
    &\hspace{0.5cm}- :C\Big[H^{(1)}(s)\Big(-\mathcal{Q}^-H^{(1)}(s)\Big) \Big]:,
\end{align}
where $C[AB]$ denotes all possible contractions between $A$ and $B$. The final equality comes from the fact
\begin{align}
    [A,B] &= :AB: + :C[AB]: \nonumber\\
    &- :BA: - :C[BA]:
\end{align}
for $A$ and $B$ given as products of creation and annihilation operators. Furthermore, it is true for all physical problems of interest because $:AB: = :BA:$, due to an equal number of fermions and antifermions, which leaves only the contractions left.
\begin{widetext}   
    \begin{align}
    \label{eq:contraction-difference}
        &:C\Big(\mathcal{G}^{(1)}(s)H^{(1)}(s)\Big): - :C\Big(H^{(1)}(s)\mathcal{G}^{(1)}(s)\Big): = \nonumber \\
        &\hspace{1cm}:C\Bigg( \int\limits_{1,2,3}4\pi\delta(\mathcal{Q}^+)f(s;\mathcal{Q}^-)r(\Lambda;\mathcal{Q}^-)\left(-\mathcal{Q^-} \right)\bar \psi(-q_1) \psi(q_2) \phi(q_3)\nonumber \\
        &\hspace{5cm}\times\int\limits_{1',2',3'}4\pi\delta(\mathcal{Q}^+)f(s;\mathcal{Q'}^-)r(\Lambda;\mathcal{Q'}^-)\bar \psi(-q_1') \psi(q_2') \phi(q_3')\Bigg):\nonumber \\
        &\hspace{1cm}-:C\Bigg( \int\limits_{1,2,3}4\pi\delta(\mathcal{Q}^+)f(s;\mathcal{Q}^-)r(\Lambda;\mathcal{Q}^-)\bar \psi(-q_1) \psi(q_2) \phi(q_3)\nonumber \\
        &\hspace{5cm}\times\int\limits_{1',2',3'}4\pi\delta(\mathcal{Q}^+)f(s;\mathcal{Q'}^-)r(\Lambda;\mathcal{Q'}^-)\left(-\mathcal{Q'}^- \right):\bar \psi(-q_1') \psi(q_2') \phi(q_3')\Bigg):\nonumber \\
        &\hspace{1cm}=-\int\limits_{\substack{1,2,3\\1',2',3'}}4\pi\delta(\mathcal{Q}^+)4\pi\delta(\mathcal{Q'}^+)f(s;\mathcal{Q}^-)f(s;\mathcal{Q'}^-)r(\Lambda;\mathcal{Q}^-)r(\Lambda;\mathcal{Q'}^-)\left(\mathcal{Q}^- -\mathcal{Q'}^- \right)\nonumber \\
        &\hspace{5cm}\times:C\Big(\bar \psi(-q_1) \psi(q_2) \phi(q_3)\bar \psi(-q_1') \psi(q_2') \phi(q_3')\Big):
    \end{align}
\end{widetext}
Equation~\ref{eq:contraction-difference} now motivates the form of $H_c(s)$, i.e. we need to include the contracted terms that arise from Eq.~\ref{eq:contraction-difference} into $H^{(2)}(s)$. To define $H_c(s$, we extract the operators and deltas from Eq.~\ref{eq:contraction-difference} only and parameterize by some arbitrary function in $s$, $F_2(s; \{q\})$, where the subscript $2$ differentiates it from the function $F_1$ in $H_I(s)$. From this, we obtain $H_c(s)$:
\begin{align}
    H_c(s) &= \int\limits_{\substack{1,2,3\\1',2',3'}}4\pi\delta(\mathcal{Q}^+)4\pi\delta(\mathcal{Q'}^+)\times\nonumber \\
    &F_2(s;\{q\})r(\Lambda, \mathcal{Q}^-)r(\Lambda, \mathcal{Q'}^-)\times\nonumber \\
    &:C\Big[\bar \psi(-q_1) \psi(q_2) \phi(q_3)\bar \psi(-q_1') \psi(q_2') \phi(q_3')\Big]:
\end{align}
The relevant non-zero field contractions are
\begin{align}
    &\wick{\c\phi(q_1) \c\phi(q_2)} = \frac{\theta(q_1)}{|q_1^+|}4\pi\delta(q_1^+ + q_2^+)\nonumber \\
    &\wick[offset=1.5em]{\c{\bar{\psi}_{i}}(q_1) \c\psi_{j}}(q_2) = -\frac{\theta(-q_1^+)}{q_1^+}4\pi\delta(q_1^+ - q_2^+)\left(\gamma_\mu q_1^\mu + m \right)_{j,i} \nonumber \\
    &\wick[offset=1.5em]{\c\psi_{i}(q_1) \c{\bar{\psi}_{j}}(q_2)} = \frac{\theta(q_1^+)}{q_1^+}4\pi\delta(q_1^+ - q_2^+)\left(\gamma_\mu q_1^\mu + m \right)_{i,j},
\end{align}
all of which will be needed when simplifying Eq.~\ref{eq:commutator-contractions}. It is now left to compute the difference in contractions in Eq.~\ref{eq:commutator-contractions}.

After grouping terms, there are two categories of equations:
\begin{equation}
\label{eq:second-order-cases}
\begin{cases}
    \frac{dH_I(s)}{ds} = \Bigg[ \Big[H_0, H_I(s) \Big], H_0\Bigg] ,\\
    \frac{dH_c(s)}{ds} = \Bigg[ \Big[H_0, H_c(s) \Big], H_0\Bigg] \\\hspace{2cm} +\Big[\mathcal{G}^{(1)}(s), H^{(1)}(s) \Big].
\end{cases}
\end{equation}
We will start by solving the first equation in~\ref{eq:second-order-cases}.

In the same way that $$\Big[H_0, H^{(1)}(s)\Big] = -\mathcal{Q}^-H^{(1)}(s)$$ was determined in Section~\ref{sec:first-order}, one can show that 
\begin{equation}
    \Big[H_0, H_I(s)\Big] = -\mathcal{Q}^-H_I(s),
\end{equation}where for this interaction vertex, $\mathcal{Q}^- = q_1^- + q_2^- + q_3^- + q_4^-$. The first equation in Eq.~\ref{eq:second-order-cases} reduces to 
\begin{equation}
    \frac{dF_1}{ds} = -\left(\mathcal{Q}^- \right)^2F_1,
\end{equation}
giving 
\begin{equation}
    F_1(s; \{q\}) = e^{-s\left(q_1^- + q_2^- + q_3^- + q_4^- \right)^2}.
\end{equation}
In terms of $\lambda$,  $H_I(\lambda)$ is:
\begin{align}
    \label{eq:HI(lambda)-appendix}
    H_I(\lambda) &= \int\limits_{1,2,3,4}4\pi\delta(\mathcal{Q}^+)f(\lambda;\mathcal{Q}^-) r(\Lambda, \mathcal{Q}^-)  \nonumber\\
    &\times :\bar \psi(-q_1) \phi(q_2)\frac{\gamma^+/2}{q_3^+ + q_4^+} \phi(q_3) \psi(q_4):,
\end{align}
where $F_1$ is replaced by the form factor.
This matches Eq.~\ref{eq:HI(lambda)}.

To solve the second equation in Eq.~\ref{eq:second-order-cases}, one must calculate the nested commutator. It can be shown that
\begin{equation}
    \Big[H_0, H_c(s) \Big] = -\Big(\mathcal{Q}^- + \mathcal{Q'}^- \Big)H_c(s).
\end{equation}
 $H_c(s)$ comes from the contractions of two first order diagrams, each of which have their own distinct $\mathcal{Q}^-$, the sum of $q^-$ going into a first order vertex. Since we have two first order vertices, we label one with a prime and one without. This implies that what we mean by $\left(\mathcal{Q}^- + \mathcal{Q'}^- \right)$ is that we sum the $q^-$'s going into both the primed and unprimed vertices, shown in Fig.~\ref{fig:contractions}.

From this, we obtain
\begin{equation}
    \Bigg[ \Big[H_0, H_c(s) \Big], H_0\Bigg] = -\Big(\mathcal{Q}^- + \mathcal{Q'}^- \Big)^2H_c(s).
\end{equation}
Now the second equation in equation~\ref{eq:second-order-cases} can be reduced to an equation for $F_2$:
\begin{align}
    \frac{dF_2}{ds} = &-\Big(\mathcal{Q}^- + \mathcal{Q'}^- \Big)^2F_2\nonumber \\
    &-f(s;\mathcal{Q}^-)f(s;\mathcal{Q'}^-)\Big(\mathcal{Q}^- - \mathcal{Q'}^-\Big).
\end{align}

A differential equation of the form $$\frac{dy}{dt} = -p(t)y(t) + q(t)$$ has the general solution $$y(t) = e^{-h(t)}\int dt e^{h(t)}q(t) + Ce^{-h(t)},$$ where $$h(t) = \int dt p(t).$$
This gives the solution:
\begin{align}
    F_2(s) = \frac12 \Big(\frac{1}{\mathcal{Q}^-}&-\frac{1}{\mathcal{Q'}^-} \Big)\Big[f(s;\mathcal{Q}^-)f(s;\mathcal{Q'}^-)  \nonumber \\ &-f(s;\mathcal{Q}^- + \mathcal{Q'}^-) \Big].
\end{align}
This gives the full form of the contracted second order terms, in terms of $\lambda$:
\begin{align}
\label{eq:Hc}
    &H_c(\lambda) = \int\limits_{\substack{1,2,3\\1',2',3'}}4\pi\delta(\mathcal{Q}^+)4\pi\delta(\mathcal{Q'}^+)\frac12\Big( \frac{1}{\mathcal{Q}^-}-\frac{1}{\mathcal{Q'}^-} \Big)\nonumber \\
    &\hspace{0.5cm}\times \Big(f(\lambda;\mathcal{Q}^-)f(\lambda;\mathcal{Q'}^-)-f(\lambda;\mathcal{Q}^- + \mathcal{Q'}^-)\Big)\nonumber\\
    &\hspace{2cm}\times r(\Lambda, \mathcal{Q}^-)r(\Lambda, \mathcal{Q'}^-) \nonumber \\
    &\hspace{0.5cm}\times :C\Big[\bar \psi(-q_1) \psi(q_2) \phi(q_3)\bar \psi(-q_1') \psi(q_2') \phi(q_3')\Big]:.
\end{align}
To get from $H_c(\lambda)$ to $H_{\text{fe}}(\lambda)$, $H_{\text{be}}(\lambda)$, and the double contraction (loop) diagrams in equation~\ref{eq:second_order}, one must compute the contractions (Eq.~\ref{eq: contractions}).
Upon calculation, one will arrive at the explicit form of the terms (not considering the counterterms) in equation~\ref{eq:second_order}.

\section{Counterterms}
\label{sec:counterterms}

Counterterms are additional terms added to the Hamiltonian to remove divergences and renormalize the theory to yield physical results. To understand the procedure of adding counterterms, it is important to start by discussing how loops arise in RGPEP solutions and by diagonalization of the Hamiltonian.

\subsection{Emergence of Loops}
Loops diagrams (corresponding to divergences) arise from the computing double contractions in $$:C\Big[\bar \psi(q_1)\psi(q_2)\phi(q_3)\bar \psi(q_1')\psi(q_2')\phi(q_3) \Big]:$$ appearing in Eq.~\ref{eq:Hc}. Each contraction contains a delta function, so the two delta functions coming from double contractions, as well as $\delta(\mathcal{Q}^+)$ and $\delta(\mathcal{Q'}^+)$ lead to $\mathcal{Q}^- + \mathcal{Q'}^- = 0$. From this, we know 
\begin{align}
    \Big(f(\lambda; \mathcal{Q}^-)f(\lambda;\mathcal{Q'}^-) - &f(\lambda;\mathcal{Q}^- + \mathcal{Q'}^-)\Big) =\nonumber \\
    &f(\lambda; \mathcal{Q}^-)f(\lambda; \mathcal{Q'}^-) - 1.
\end{align}
Without considering the regulating function, the existence of the $-1$ is the cause of the divergence because there is no form factor to weaken the interaction and the integral will diverge. The regulating functions remove the divergence, but now the mass spectrum will be sensative to the choice of $\Lambda$. Renormalization (and the choice of counterterms) is done to remove this sensitivity while retaining the finite results that $\Lambda$ provides. 





\subsection{Regularization}
The first step of renormalization is always regularization. Two common regularization techniques used in standard field theory calculations in a Lagrangian approach are Pauli-Villars~\cite{PV-regularization} and dimensional regularization~\cite{dim-reg}.

In our approach, regularization occurs by placing a cutoff $\Lambda$ on the total \textit{difference} between energy in minus energy out of a given interaction vertex. We ultimately want this regulator $\Lambda \rightarrow \infty$, but different finite $\Lambda$'s will give different results. A counterterm is chosen such that the need for a regulator is removed. 

The full form of the regulator is taken to be the same form as the form factor such that the regulating function $r(\Lambda, \mathcal{Q}^-) = e^{- \left(\mathcal{Q}^-/\Lambda \right)^2}$ modifying the interaction vertices leads to all effective interaction vertices now being modified by $e^{- \left(\mathcal{Q}^- \right)^2\Big(1/\lambda^2 + 1/\Lambda^2\Big)}$.

\subsection{Perturbative Mass Renormalization}
In diagonalizing the bare Hamiltonian, higher order terms appear.
Perturbation theory allows one to understand the higher order terms, and fix counterterms to eliminate divergent terms.
We begin with the definition of the perturbative correction the $n^{th}$ eigenvalue:

\begin{align}
\label{eq:m_correction}
    E_n = E_n^{(0)} &+ \langle n^{(0)}|V|n^{(0)}\rangle \nonumber \\
    & +\sum_{k \neq n}\frac{|\langle k^{(0)}|V|n^{(0)}\rangle|^2}{E_n^{(0)} - E_k^{(0)}} + \mathcal{O}\left(g^3\right)
\end{align}

We define the \textit{renormalized fermion mass} $m^2$ as:

\begin{equation}
    m^2(g) \equiv M^2(\lambda, g)\Bigr|_{Q = 1},
\end{equation}
i.e. the eigenvalue corresponding to the state in the $Q = 1$ sector that is dominated by a single fermion (this will be the ground state of this sector). 

In order to properly renormalize a parameter, a \textit{renormalization condition} must be set. The condition for the mass term is 
\begin{equation}
    m^2(g) = m^2,
\end{equation}
i.e. the eigenvalue of the $Q = 1$ sector of the effective Hamiltonian is the physical mass $m^2$. 




With this definition, the perturbative corrections, up to $\mathcal{O}\left(g^2\right)$, to this eigenvalue is: 

\begin{equation}
    \frac{m^2}{P^+} = \Big(\frac{m^2}{P^+}\Big)^{(0)}+\Big(\frac{m^2}{P^+}\Big)^{(1)}+\Big(\frac{m^2}{P^+}\Big)^{(2)}.
\end{equation}
$\Big(m^2/P^+\Big)^{(0)}$ is the unperturbed eigenvalue, i.e the eigenvalue of the canonical Yukawa Hamiltonian with $g = 0$.
This is simply $m^2$, the parameter in the Lagrangian. Furthermore,

\begin{align*}
    \Big(\frac{m^2}{P^+}\Big)^{(1)} &= \langle f| H^{(1)}|f\rangle=0,
\end{align*}
since $H^{(1)}|f\rangle \sim |fb\rangle$.

There are two nonzero contributions to the second order correction. 
The first comes from the standard second order correction in Eq.~\ref{eq:m_correction}, that is the perturbative loops arising from two first order effective diagrams. 
A correction also arises from the loop diagram $H_{\delta m^2}(\lambda)$ coming from solving the RGPEP equation, i.e. Eq.~\ref{eq:fermion-loop}. 

\begin{align*}
    \Big(\frac{m^2}{P^+}\Big)^{(2)} &= \langle f|H_{\delta m^2}(\lambda) |f\rangle\nonumber\\
    &+ \langle f|H^{(1)}(\lambda) \frac{1}{m^2/P^+ - H_0}H^{(1)}(\lambda)|f\rangle.
\end{align*}
The second term comes from the $\mathcal{O}(g^2)$ term in $E_n$, Eq.~\ref{eq:m_correction}. 
This is clear by writing the resolvent, $(m^2/P^+ - H_0)^{-1}$ in its spectral decomposition \cite{brodsky1998quantum} and then computing its expectation value with $H^{(1)}(\lambda)\ket{f}$.

For explicit calculations see~\cite{serafin-yukawa}. We give compact forms here: 

\begin{align}
    &\langle f|H_{\delta m^2}(\lambda) |f\rangle  = I_F\Big(\lambda^{-2} + \Lambda^{-2} \Big) - I_F\Big(\Lambda^{-2}\Big) \\
    \label{eq:perturbative-loop}
    &\langle f|H^{(1)}(\lambda) \frac{1}{m^2/P^+ - H_0}H^{(1)}(\lambda)|f\rangle \nonumber \\
    &\hspace{4cm}= -I_F\Big(\lambda^{-2} + \Lambda^{-2} \Big),
\end{align}
for $P^+ = 1$, where
\begin{align}
    I_F(s) &= g^2 \int_0^1\frac{dx}{4\pi x(x-1)}\nonumber\\
    &\times\Big( \frac{m^2}{x} + \frac{\mu^2}{1-x} - m^2\Big)^{-1}\frac{m^2(1 + x)^2}{x}\nonumber \\
    &\times \exp\left({-\frac{2s}{\left( q^+\right)^2}\left(\frac{m^2}{x} + \frac{\mu^2}{1-x} - m^2\right)^2}\right).
\end{align}
$I_F$ is divergent as $\Lambda \rightarrow \infty$, but finite for $\Lambda < \infty$.

Taking both contributions into account,
\begin{equation}
    ({m^2})^{(2)} = - I_F\Big(\Lambda^{-2}\Big).
\end{equation}
This gives the form of the perturbative correction to $m^2$:
\begin{equation}
\label{eq:second-order-mass}
    m^2 = m^2 -I_F\Big(\Lambda^{-2}\Big)+ \mathcal{O}\Big(g^3 \Big).
\end{equation}

Eq.~\ref{eq:second-order-mass} explicitly gives the proper form of the fermion mass counterterm that needs to be added to $H(\lambda)$ to remove the divergence: 

\begin{equation}
    \label{eq:fermion-mass-counterterm}
    X_{\delta m^2} \sim I_F\Big(\Lambda^{-2}\Big).
\end{equation}
The combined contribution of $H_{\delta m^2}$ and $X_{\delta m^2}$ in $H(\lambda)$ leads to a term $$\sim I_F\Big(\lambda^{-2} + \Lambda^{-2} \Big) b^\dagger b$$ in the effective Hamiltonian.
The regulator can now be removed $\Lambda \rightarrow \infty$ with results becoming finite.

The implicit loop diagram arising from diagonalization goes as $$\sim -I_F\Big(\lambda^{-2}\Big)b^\dagger b$$ (see Eq.~\ref{eq:perturbative-loop}).
The explicit loop coming from $$H_{\delta m^2} + X_{\delta m^2}\sim I_F\Big(\lambda^{-2}\Big)b^\dagger b.$$
These two (implicit and explicit) loops now \textit{cancel} which leads to the renormalization condition $m^2 = m^2$ being met. This can be verified in the numerical results in Fig.~\ref{fig:renormalized}.


The boson renormalization is slightly different. 
We define 
\begin{equation}
    \mu^2(g) \equiv M^2(\lambda, g) \Bigr|_{Q = 0},
\end{equation}
with renormalization condition 
\begin{equation}
    \mu^2(g) = \mu^2(g = 0).
\end{equation}
One can work out the same steps to get the boson mass counterterm: 

\begin{equation}
    X_{\delta \mu^2} = I_B\Big(\Lambda^{-2} \Big),
\end{equation}
where 

\begin{align}
    I_B(s) &= 2g^2 \int_0^1\frac{dx}{4\pi x(x-1)} \nonumber\\
    &\Big(\frac{m^2}{x} + \frac{m^2}{1-x} - \mu^2\Big)^{-1}\frac{m^2(2x-1)^2}{x(1-x)}\nonumber \\
    &\times \exp\left({-\frac{2s}{\left(q^+\right)^2}\left(\frac{m^2}{x} + \frac{m^2}{1-x} - \mu^2\right)^2}\right).
\end{align}
The effective Hamiltonian gets a term $\sim I_B\Big(\lambda^{-2} \Big)a^\dagger a$.

\end{document}